\documentclass[journal=cmatex,manuscript=article]{achemso}

\pdfoutput=1
\mciteErrorOnUnknownfalse 

\usepackage[version=3]{mhchem} 
\usepackage{xcolor}
\usepackage[english]{babel}
\usepackage{amsmath}
\usepackage{bm}
\usepackage{graphicx}
\usepackage[colorlinks=true, allcolors=blue]{hyperref}
\usepackage{gensymb}
\usepackage[section]{placeins}

\SectionNumbersOn

\author{Pieter Dobbelaere}
\affiliation[UGENT]
{Center for Molecular Modeling
Ghent University
Tech Lane Ghent Science Park Campus A, 9052 Zwijnaarde, Belgium}

\author{Sander Vandenhaute}
\affiliation[UGENT]
{Center for Molecular Modeling
Ghent University
Tech Lane Ghent Science Park Campus A, 9052 Zwijnaarde, Belgium}

\author{Veronique Van Speybroeck}
\affiliation[UGENT]
{Center for Molecular Modeling
	Ghent University
	Tech Lane Ghent Science Park Campus A, 9052 Zwijnaarde, Belgium}
\email{veronique.vanspeybroeck@ugent.be}

\title[]
  {Cluster-based machine learning potentials to describe disordered metal-organic frameworks up to the mesoscale}

\begin{document}



\begin{abstract}
Metal-organic frameworks (MOFs) are highly interesting and tunable materials.
By incorporating spatial defects into their atomic structure, MOFs can be finetuned to exhibit precise chemical functionalities, extending their applicability in various technological fields.
Defect engineering requires a fundamental understanding of the nature of spatial disorder and consequent changes in material properties, which is currently lacking.
We introduce the cluster-based learning methodology, enabling the development of state-of-the-art machine learning potentials (MLPs) from defective systems at any length scale.
Our method identifies atomic interactions in bulk structures and extracts local environments as finite molecular fragments to augment the model's training data where needed.
We show that cluster-based learning delivers MLPs capable of accurately describing spatial defects in mesoscopic systems with over twenty thousand atoms.
Afterwards, we select our best model to investigate some major mechanical properties of spatially disordered UiO-66-derived structures, elucidating the influence of defect concentration and composition on material behaviour.
Our analysis includes large supercell structures, demonstrating that (near-) \textit{ab initio} accuracy is within reach at the mesoscale.
\end{abstract}


\section{Introduction} \label{sec_intro}

Metal-organic frameworks, MOFs, are porous crystalline solids that have evolved into versatile materials with many technological and industrial applications in e.g., heterogeneous catalysis, gas sorption and separation or nanoscopic actuating and sensing. \cite{ref_lei_design_2014,ref_burtch_mechanical_2018,ref_rogge_metalorganic_2017,ref_temmerman_computational_2024}
Structurally, MOFs comprise a topological lattice made from several secondary building units, namely metal nodes and organic ligands. \cite{ref_yaghi_reticular_2003,ref_furukawa_chemistry_2013}
In computational analyses, they are usually treated as well-ordered and pristine molecular systems.
However, many recent studies have highlighted that the strength of MOFs lies in their ability to encapsulate atomically precise functions through defects. \cite{ref_bennett_interplay_2017,ref_dissegna_defective_2018}
Many enticing MOF properties are heavily influenced or modulated by inhomogeneities in the perfect framework.
Such deviations from order exist in every realistic structure, appearing in different forms and over multiple length scales.
We find point defects like metal atom substitutions or missing ligand vacancies on the nanoscale, whereas mesoscopic disorder materialises as larger cavities or mesopores, regions of phase coexistence and surface boundaries in finite crystals. \cite{ref_cheetham_defects_2016, ref_dai_highly_2024}
Understanding how various types and arrangements of spatial disorder enhance or interfere with desired material characteristics, is crucial to exploit this configurational freedom and tailor frameworks to their intended application.
\cite{ref_feng_tailoring_2019,ref_xiang_synthesis_2020}
To unlock the full potential of MOFs through defect engineering, we require computationally efficient and accurate modelling techniques that can describe spatial disorder up to the mesoscale. \\

Over the last few years, advances in machine learning potentials (MLPs) have initiated a new era for molecular modelling of functional materials. \cite{ref_behler_perspective_2016,ref_mueller_machine_2020,ref_behler_machine_2021,ref_friederich_machine_learned_2021, ref_morrow_how_2023, ref_domina_cluster_2023}
These cutting-edge neural networks parametrise a molecular potential energy surface (PES) by learning atomic interactions from underlying quantum mechanical (QM) calculations.
They can accurately reproduce their reference level of theory (LOT) at a (comparatively) vanishingly small inference cost, once trained.
In simulations, MLPs assume the role of interatomic potential, similar to classical force fields - albeit much more faithful to QM behaviour and without enforcing a fixed bonding topology. \cite{ref_jia_pushing_2020,ref_xie_ultra_fast_2023}
Their main weakness is the notoriously poor ability to describe out-of-dataset structures; a phenomenon aptly named the extrapolation problem. \cite{ref_kosanovich_improving_1996,ref_mishin_machine_learning_2021, ref_mahmoud_assessing_2025}
Therefore, the accuracy and transferability of any MLP depend vitally on the chemical and configurational space covered by its training data.
Developing capable potentials for disordered frameworks entails constructing representative datasets, which becomes computationally intractable at the length scales needed to represent disorder.
At present, the main hurdle holding back MLP development for defective materials is the cost of \textit{ab initio} data generation. \\

To achieve our goal, we introduce the chemical environment of an atom as the key concept governing molecular interactions. Intuitively, a chemical environment can be understood as the sphere of influence between an atom and its periphery, uniquely encompassing all non-negligible interactions with neighbouring atoms and externally applied fields. The principle of nearsightedness of electronic matter - which states that a perturbing potential only causes a finite change in the local electron density, whose magnitude decreases monotonically with increasing distance to said perturbation
\cite{ref_kohn_nearsightedness_2005} -
presumes a finite radius of electronic interaction and lies at the core of many linearly scaling density functional theory (DFT) implementations.
\cite{ref_ratcliff_flexibilities_2020,ref_prentice_span_2020}
Consequently, environments must have a limited spatial extent and can be treated as local molecular properties. Hence, a dataset of QM calculations can also be interpreted as a collection of independent environment-interaction pairs. If the dataset contains all environments to fully describe every atomic interaction present in a given system of interest, we say it is representative of that system. \\

\begin{figure}
    \centering
    \includegraphics[width=\linewidth]{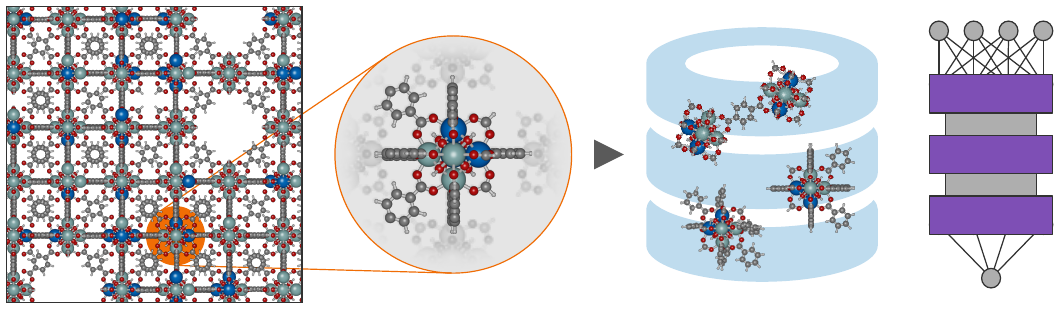}
    \caption{Cluster-based learning. Local chemical environments in bulk frameworks are extracted as finite clusters to capture specific atomic interactions, enabling MLPs to learn spatial disorder at the mesoscale.}
    \label{fig_intro}
\end{figure}

In this work, we propose that representative datasets can be constructed for arbitrarily large (and disordered) structures by extracting individual environments in the form of small molecular fragments (see Figure \ref{fig_intro}).
The procedure involves two steps: (i) identify unknown chemical environments using a learned MLP characterisation and (ii) separate those environments from the molecular bulk as isolated fragments.
This `cluster-based learning' idea circumvents computational bottlenecks and allows us to selectively design the chemical and configurational space spanned by the training data. \\

For optimal efficiency, we embed our methodology into an active learning (AL) scheme - a machine-learning paradigm where models are iteratively refined by cycling between data acquisition and retraining phases. \cite{ref_settles_al_2009,ref_jinnouchi_al_2020, ref_young_transferable_2021, ref_sharma_quantum_accurate_2024}
In a molecular modelling context, AL foregoes any expensive \textit{ab initio} sampling conventionally used to generate reference data, making it exceptionally attractive. \cite{ref_ang_active_2021}
We recently developed an in-house AL framework, Psiflow \cite{ref_vandenhaute_machine_2023},
that automates MLP training for periodic systems and molecules.
Here, we implement a novel cluster-based data acquisition algorithm and provide an interface with existing Psiflow functionality.
Two interesting use cases are tackled in this work:

\begin{itemize}
    \item \underline{Large structures:} molecular systems are often spatially complex and can only be described in large unit cells. This is especially true in the world of MOFs, e.g., the isoreticular DUT-series (1000-2000 atoms) \cite{ref_ying_pressure_induced_2021}, MIL-100 ($\sim$ 3000 atoms) \cite{ref_dona_cost_effective_2020} or plenty of other hierarchically porous frameworks \cite{ref_feng_hierarchically_2020}. To also include defects and disorder, structures must grow even larger. With so many degrees of freedom, \textit{ab initio} evaluations become downright impossible. Deconstructing large cells into smaller fragments decouples their characteristic dimensions from the cost of QM computations. Provided we can define a suitable partitioning into compact clusters, this strategy enables the construction of accurate MLPs for any atomic structure regardless of its inherent size.

    \item \underline{Transferable MLPs:} selectively filtering for missing environments can efficiently increase model transferability while reducing data redundancy. Consider, for example, the prototypical MOF UiO-66 and its isoreticular cousin UiO-67. To train an MLP, one could naively collect periodic data for both systems separately. This is wasteful, as both frameworks share most atomic interactions. Instead, we could construct a dataset for UiO-66 first, and later extend it with fragments that contain the unique environments of UiO-67. Gathering data incrementally eliminates unnecessary QM evaluations and is especially suited to developing universal models that describe large families of materials.

\end{itemize}

We should concede one caveat: since our idea involves finite fragments for training, models will only learn short-range interactions.
However, reference data can always contain mixed boundary conditions, i.e., some periodic structures - describing long-range phenomena - and molecular clusters - containing local environments.
When electrostatics or electron dispersion become dominant energy contributions, MLPs are often enhanced by including physical priors or predicting partial charges, for example. \cite{ref_anstine_machine_2023}
These corrections remain applicable in conjunction with the proposed approach. \\

In this work, we explain the inner workings of the cluster-based learning procedure; how chemical environments are identified based on an internal MLP representation, how clusters are designed to extract specific interactions and how everything fits into an automated AL workflow.
Using UiO-66 as a case study, we investigate how small point defects alter framework interactions in pristine MOFs, show when trained MLPs fail and how to correct spurious behaviours by extending existing datasets.
To highlight the general applicability of our methodology, we successfully learn disordered systems at diverse length scales, containing different types and concentrations of spacial disorder (see Figure \ref{fig_intro}).
Finally, with a robust and transferable MLP at hand, we explore the pressure response of various systems in the UiO-66 family and uncover fundamental relations between induced disorder and mechanical resilience in MOFs.

\section{Methods} \label{sec_methods}

Below, we discuss the two major components that form our cluster-based learning implementation: uncertainty quantification and cluster extraction (see Figure \ref{fig_al}.A).
First, we characterise chemical environments in sample structures using a learned MLP representation and quantify model uncertainty - a measure of confidence in its predictive accuracy.
Excessive uncertainties indicate MLP extrapolation and unreliable inference, showing deficiencies in the training dataset.
Then, we extract regions of high uncertainty from the bulk material into representative molecular fragments, using an algorithm to design clusters that mimic environments found in the original system.
This data acquisition method is automated to easily enable \textit{ab initio} evaluation of bulk interactions.
It can deliver state-of-the-art MLPs for arbitrarily large structures at very modest computational expense, without requiring manual intervention. \\

We formalise the intuitive definition of a chemical environment in Section \ref{sec_chem_env}.
A brief overview of the AL workflow is given in Section \ref{sec_al}.
Mathematical details on MLP uncertainty quantification are provided in Section \ref{sec_uncertainty}.
Finally, Section \ref{sec_extraction} discusses the intricacies of the cluster extraction procedure.

\begin{table}[]
\centering
\begin{tabular}{lll}
\hline
$\epsilon$ && chemical environment  \\
$S$ && molecular system \\
$D$ && dataset of structures \\
$\bm{F}$ && atomic force vector \\
$\bm{\mathcal{F}}$ && atomic feature vector \\
\hline
\end{tabular}
\caption{Summary of used symbols.}
\label{table_cheat_sheet}
\end{table}

\begin{table}[]
\centering
\begin{tabular}{lll}
\hline
pr && pristine or defect-free  \\
ld && linker defect \\
hf && hafnium substitution \\
reo && reo node defect \\
\hline
\end{tabular}
\caption{Summary of used abbreviations.}
\label{table_abbreviation}
\end{table}

\subsection{Chemical environments} \label{sec_chem_env}

Within a molecular structure, we define the chemical environment $\epsilon_i$ of atom $i$ as the radius of interaction between this atom and its neighbourhood.
It encompasses everything the central atom can `feel', such as surrounding atoms and externally applied fields, and is necessarily local by the principle of nearsightedness of electronic matter. \cite{ref_kohn_nearsightedness_2005}
Concretely, an environment $\epsilon_i$ consists of all (structural) information to uniquely describe the total sum of perceptible influences on atom $i$.
We will drop the subscript when referring to any (generic) environment. \\

Two isolated structures connected by rotations, translations and inversions are physically identical; they carry the same atomic interactions, although the resulting forces could differ in orientation.
Accordingly, these symmetries will project $\epsilon$ onto itself.
We say that $\epsilon$ is invariant for transformations of the Euclidean group E(3), abstracting away directional degrees of freedom.
This is analogous to an invariant molecular energy giving rise to equivariant atomic forces, which transform like vectors under E(3). \\

Simplifying further notation (see Table \ref{table_cheat_sheet}), we denote a molecular system with $S$, a dataset of structures with $D$, and use $D(\epsilon)$ to explicitly refer to the $\epsilon$ contained within $D$.
When sampling in some thermodynamic ensemble, system $S$ has access to a volume of its configuration space in accordance with state variables ENS (e.g., the NPT ensemble with P = 1 bar and T = 300 K).
Consequently, $S$ can occupy a related volume in `chemical environment space'. $S(\epsilon)|_{\text{ENS}}$ represents all $\epsilon$ that appear in any configuration of $S$ under thermodynamic conditions ENS, as schematically shown in Figure \ref{fig_al}.B.
Adopting these conventions, we propose:

\begin{equation}
D \text{ is representative for } S \text{ under ENS } \Longleftrightarrow \forall \epsilon_i \in S(\epsilon)|_{\text{ENS}}, \exists \epsilon_j \in D(\epsilon): \epsilon_i \approx \epsilon_j
\label{eq_repr}
\end{equation}

where `$\approx$' will become meaningful later (see Section \ref{sec_force_match} and Section \ref{sec_feature_match}). \\

Computationally, we can describe molecular structures using various LOTs. The properties of every $\epsilon$ - i.e., its shape and spatial extent or the relative importance of different interactions - will depend on the chosen computational method.
Therefore, we introduce a further specification, $\epsilon^{\text{lot}}$, to distinguish the LOT employed.
In this work, reference datasets are constructed using DFT and all simulations rely on MLP inference.
Correspondingly, we consider both $\epsilon^{\text{dft}}$ and $\epsilon^{\text{mlp}}$ as approximations to the true QM $\epsilon$.
In practice, the superscript `dft' refers to a specific set of computational parameters (functional, basis set, etc.) and `mlp' points to a particular MLP. \\

The fundamental ansatz of cluster-based learning is the idea that $\epsilon$ from bulk structures can be captured and extracted in finite molecular fragments. Atom $i$ should experience the same total interaction in the designed cluster and the original parent system, for all relevant LOTs:

\begin{equation}
\left( \epsilon^{\text{dft}}_i, \epsilon^{\text{mlp}}_i \right)_{\text{bulk}} = \left( \epsilon^{\text{dft}}_i, \epsilon^{\text{mlp}}_i \right)_{\text{cluster}}
\label{eq_env_match}
\end{equation}

Equation \ref{eq_env_match} represents the condition of `environment matching', and is a prerequisite for MLPs to learn bulk interactions from a dataset of clusters.
We can enforce it by defining methods to characterise and compare different $\epsilon$, namely force matching (Section \ref{sec_force_match}) and feature matching (Section \ref{sec_feature_match}). The latter requires a quantitative representation of $\epsilon$, while the former only involves evaluated force labels.

\subsubsection{Force matching} \label{sec_force_match}

Atomic forces are local observables that directly reflect underlying molecular interactions.
Disregarding orientation, equal interactions will cause identical forces.
If the original structure and extracted cluster are aligned so that corresponding atoms overlap perfectly, forces obey:

\begin{equation}
\left( \epsilon_i \right)_{\text{bulk}} = \left( \epsilon_i \right)_{\text{cluster}} \Rightarrow \left( \bm{F}_i \right)_{\text{bulk}} = \left( \bm{F}_i \right)_{\text{cluster}}
\label{eq_force_match}
\end{equation}

where $\bm{F}_i$ is the total force vector felt by atom $i$.
Equation \ref{eq_force_match} is not injective; many $\epsilon$ give rise to the same $\bm{F}$.
Nevertheless, we will assume the inverse holds too.
In other words: if the force on atom $i$ matches between fragment and parent, their environments should be equivalent. \\

Force matching prescribes a transparent algorithm to find appropriate clusters:
(i) evaluate $(\bm{F}_i)_{\text{bulk}}$, (ii) evaluate $(\bm{F}_i)_{\text{cluster}}$ for a series of candidate fragments and (iii) find the closest match between (i) and (ii), as that cluster encloses the best approximation of $(\epsilon_i)_{\text{bulk}}$.
At large length scales, (\textit{ab initio}) evaluations of the parent structure might no longer be possible.
As a workaround, one can extrapolate the evolution of $(\bm{F}_i)_{\text{cluster}}$ for fragments of increasing size to estimate a limit for $(\bm{F}_i)_{\text{bulk}}$.
An example of force matching for $\epsilon^{\text{dft}}$ and $\epsilon^{\text{mlp}}$ is discussed in Section SI.6.

\subsubsection{Feature matching} \label{sec_feature_match}

During training, neural network MLPs learn to encode the surroundings of an atom into a descriptive feature representation, progressively increasing the level of abstraction throughout several hidden layers. The range of $\epsilon^{\text{mlp}}$ is limited by the atomic interaction radius $r_{\text{max}}$ of the model, determining how far it can `look ahead'.
This is only an upper bound, as $r_{\text{max}}$ can be chosen arbitrarily large, whereas the intrinsic size of $\epsilon^{\text{mlp}}$ cannot be. To accurately infer molecular interactions, the MLP must discriminate various $\epsilon^{\text{mlp}}$ by its features. We define a feature descriptor $\bm{\mathcal{F}}$ constructed from (a subset of) $n$ invariant network nodes - e.g., the final hidden layer - to identify all $\epsilon^{\text{mlp}}$. These $n$-dimensional $\bm{\mathcal{F}}$-vectors span feature space. In analogy to Equation \ref{eq_force_match}, the relation between environment and descriptor is given by:

\begin{equation}
\left( \epsilon^{\text{mlp}}_i \right)_{\text{bulk}} = \left(\epsilon^{\text{mlp}}_i \right)_{\text{cluster}} \Rightarrow \left( \bm{\mathcal{F}}_i \right)_{\text{bulk}} = \left( \bm{\mathcal{F}}_i \right)_{\text{cluster}}
\label{eq_feature_match}
\end{equation}

If the MLP architecture and chosen $\bm{\mathcal{F}}$ are sufficiently expressive, the inverse of Equation \ref{eq_feature_match} will also hold.
Therefore, we can identify $\epsilon^{\text{mlp}}_i$ with a point in feature space and associate differences in $\bm{\mathcal{F}}$ with a degree of (dis)similarity between $\epsilon^{\text{mlp}}$.
Note that comparing multiple $\bm{\mathcal{F}}$ is only meaningful for a single MLP, because altering network weights changes the structure of feature space.
In Section \ref{sec_results_feature_space}, we show that a trained model does indeed distinguish distinct atomic interactions internally, i.e., that $\bm{\mathcal{F}}$-vectors embed chemical information.

\subsection{Active Learning} \label{sec_al}

\begin{figure}
\centering
\includegraphics[width=0.6\linewidth]{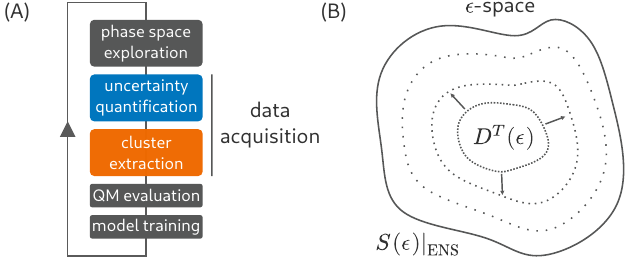}
\caption{(A) Schematic overview of the active learning workflow. (B) Every cycle $D^T(\epsilon)$ grows in $\epsilon$-space, incrementally approaching the target  $S(\epsilon)|_{\text{ENS}}$.}
\label{fig_al}
\end{figure}

We aim to create an accurate MLP for system $S$ under thermodynamic conditions ENS while minimising computational costs and linear execution time.
AL workflows achieve this objective by iteratively expanding a training set $D^T$ (superscript T for training) to be more representative of $S(\epsilon)|_{\text{ENS}}$ until Equation \ref{eq_repr} is fulfilled.
The process - coined an AL campaign - consists of several AL cycles and is illustrated in Figure \ref{fig_al}.A. \\

Following a one-off initialisation step, every cycle involves phase space exploration, data acquisition and evaluation, and model retraining stages.
During exploration, the goal is to sample new $\epsilon$ from $S(\epsilon)|_{\text{ENS}}$ by probing additional configurations of $S$, using a model trained in the previous cycle.
The resulting structures are analysed for unknown $\epsilon \notin D^T(\epsilon)$, which are extracted into finite clusters and evaluated at an \textit{ab initio} LOT.
These fragments are appended to the existing dataset, extending the region of $\epsilon$-space described by $D^T(\epsilon)$ (see Figure \ref{fig_al}.B).
Finally, the MLP is retrained with a newly improved dataset, concluding the current cycle.
An AL campaign continues for a predefined number of iterations or until some accuracy threshold has been met.
Section SI.4 provides concrete descriptions of a full cycle as implemented in this work.

\subsection{Uncertainty quantification} \label{sec_uncertainty}

The computational efficiency of AL workflows can be drastically improved through active (as opposed to random) sample selection.
These techniques aim to maximise transferability and reduce data redundancy of $D^T$ by identifying out-of-dataset structures.
Broadly, they assess MLP uncertainty - whether we expect the model to reproduce atomic interactions correctly - for unlabelled sample configurations.
High uncertainty suggests poor inference accuracy and the presence of unfamiliar environments, making it worthwhile to incorporate those $\epsilon$ into $D^T(\epsilon)$. \\

We find a myriad of methods to quantify uncertainty in recent AL literature.
Effective estimates for cluster-based learning should not involve \textit{ab initio} calculations, only rely on the current model and dataset and provide per-atom predictions.
If GPU compute time abounds, query-by-committee approaches have proven general and widely successful in machine learning. \cite{ref_zhang_active_2019,ref_smith_less_2018}
Gaussian Process models can immediately leverage their built-in predictive uncertainty. \cite{ref_zhai_active_2020,ref_vandermause_fly_2020}
Other data-driven implementations employ statistical measures computed over kernel or feature embeddings of $D^T$. \cite{ref_angelov_model_free_2017,ref_zaverkin_exploring_2022,ref_tan_enhanced_2024,ref_zhu_fast_2023} \\

\begin{figure}[]
\centering
\includegraphics[width=\linewidth]{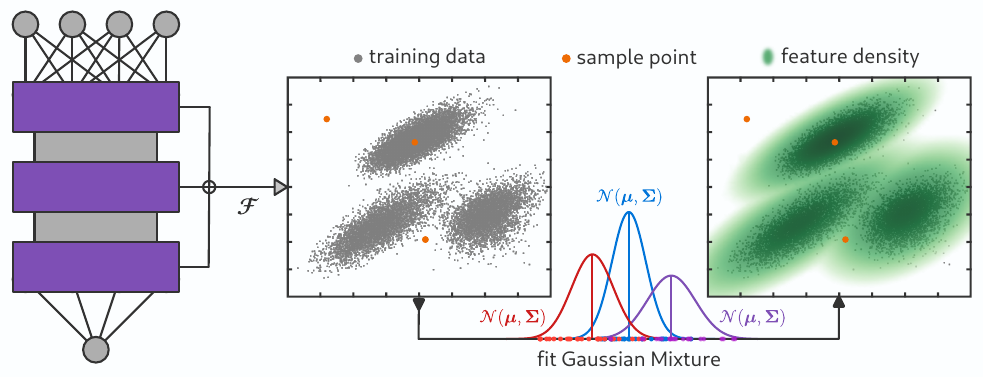}
\caption{Descriptor $\bm{\mathcal{F}}$ casts $D^T(\epsilon)$ into feature space. A GMM is fit to represent the density of training points. For any sample $\epsilon$ (orange), the density likelihood represents similarity with $D^T(\epsilon)$. In trained MLPs, this likelihood is inversely correlated with model uncertainty.}
\label{fig_dens_fit}
\end{figure}

Here, we take the latter route, using a learned descriptor $\bm{\mathcal{F}}$ to convert the abstract $D^T(\epsilon)$ into a concrete set of vectors that describe all interactions of $D^T$ in MLP feature space.
After fitting these features with a density distribution, the likelihood of $\bm{\mathcal{F}}_i$ estimates how similar  $\epsilon^{\text{mlp}}_i$ is to existing training data.
Low likelihoods imply few nearby data points, indicating out-of-dataset $\epsilon$, or vice-versa (see Figure \ref{fig_dens_fit}).
Following model training, we assume that MLP uncertainty and density likelihoods are inversely correlated (see Section SI.7). \\

Many unsupervised density models are suited for this task.
We opt for a Gaussian Mixture Model (GMM) \cite{ref_moitra_2018}, which is a weighted summation of multivariate Gaussians:

\begin{equation}
\text{GMM}\left[ D^T(\epsilon), \bm{\mathcal{F}} \right] = \sum_m^M a_m \mathcal{N}_n \left( \bm{\mu}_m, \bm{\Sigma}_m \right),
\end{equation}

where $n$ is the dimension of $\bm{\mathcal{F}}$, $M$ denotes the number of mixture components, $a_m$ is a relative weight factor for each component and $\mathcal{N}_n (\bm{\mu}, \bm{\Sigma})$ is an $n$-dimensional normal distribution with mean vector $\bm{\mu}$ and covariance matrix $\bm{\Sigma}$.
The total distribution is normalised to unity.
We utilise the Bayesian information criterion to decide an appropriate value for $M$. \cite{ref_stoica_model_order_2004} \\

If the curse of dimensionality \cite{ref_altman_curses_2018} complicates density construction, one can resort to techniques such as principal component analysis (PCA) \cite{ref_jolliffe_principal_2016}
to reduce the dimension of $\bm{\mathcal{F}}$ to a more manageable $n'$, whilst retaining most information.
This is also useful in visualisations.
For increased model flexibility (see Section \ref{sec_extraction}), we group $\bm{\mathcal{F}}_i$ according to the atomic number of atom $i$ and build independent distributions for every element in $D^T$.
If no training data is available for a particular element, matching $\bm{\mathcal{F}}$ are given an artificial likelihood of zero.

\subsection{Cluster extraction} \label{sec_extraction}

Molecular fragments are finite clusters designed to encapsulate some $\epsilon$ from a larger parent structure.
Using the uncertainty approach of Section \ref{sec_uncertainty}, we can scan sample configurations for unknown $\epsilon \notin D^T(\epsilon)$, effectively creating a heatmap of where new interactions are located (see Figure SI.3).
If we find a spatially concentrated group of $\epsilon$ with high uncertainty, they should be extracted into a new fragment. \\

The first component of a cluster under design is its core, containing all atoms whose $\epsilon$ are to be added to $D^T(\epsilon)$.
To fulfil Equation \ref{eq_env_match} (environment matching), we must envelop the chosen core in a suitable mantle of atoms from the original structure, serving as padding to mimic bulk environments.
This is achieved using the force matching approach (see Section \ref{sec_force_match}).
At this point, our cluster successfully captures the core $\epsilon$, but includes several dangling bonds at its surface, complicating future \textit{ab initio} evaluations.
We form a termination layer to saturate broken bonds - creating new atoms not in the parent system - and form a chemically valid, preferably closed-shell and charge-neutral fragment.
In total, the cluster consists of three distinct regions: (i) a `core' realising Equation \ref{eq_env_match}, (ii) a `mantle' violating Equation \ref{eq_env_match} and (iii) a termination layer without a counterpart in the original structure (see Figure SI.2). \\

This section only sketches a rough outline, rather than providing a concrete extraction recipe.
Defining reasonable clusters requires chemical insight, and depends strongly on the material of interest.
We seek to capture a maximal amount of information in each fragment, whilst keeping computational costs to a minimum.
A more thorough discussion of this process, applied to MOFs, is given in Section SI.5.
We provide an in-depth example of cluster design through force matching in Section SI.6.

\section{Computational details} \label{sec_comp_det}

We present a concise overview of the main computational choices made in this work, and will regularly refer to the SI for more exhaustive explanations. \\

\underline{QM evaluations:} every \textit{ab initio} evaluation is performed by the GPAW software engine \cite{ref_mortensen_2005}, because it allows both finite and periodic boundary conditions.
This way, clusters and periodic structures are evaluated in a consistent way.
We work with its finite difference grid formulation, employ the Perdew–Burke-Ernzerhof (PBE) DFT functional \cite{ref_perdew_generalized_1996} with Becke-Johnson D3 dispersion correction \cite{ref_johnson_post_hf_2006} and use a basis grid spacing of 0.175 \AA.
Further details can be found in Section SI.1. \\

\underline{MLPs:} we choose NequIP \cite{ref_batzner_nequip_2022} as the fundamental architecture for all MLPs, activating equivariant features by setting the rotation order $l > 0$. In Section \ref{sec_results}, model accuracy is used to assess the quality of $D^T$.
Because inference flaws should directly reflect dataset deficiencies, we allow MLPs to reach optimal performance and extract maximal information - in practice, until their validation error stops decreasing.
A comprehensive specification of network hyperparameters and training setup is given in Section SI.2.
Note that our methodology is architecture-agnostic, and other MLP frameworks could be used as well.\\

\underline{Simulations:} molecular dynamics (MD) is performed with either the OpenMM engine \cite{ref_eastman_openmm_2017} or the in-house YAFF software \cite{ref_verstraelen_yaff}. OpenMM offers more efficient simulation algorithms but is limited in functionality for periodic systems, whereas YAFF implements many ensembles specifically geared towards periodic boundary conditions. Simulation parameters can be found in Section SI.3. \\

\underline{Dataset generation:} Periodic datasets for system $S$ are generated from OpenMM MD in the isobaric-isothermal (NPT) ensemble at 600 K and various pressures, ensuring diversity of $\epsilon$.
Structures are selected uniformly within a fixed volume range around the equilibrium volume of $S$.
However, meticulously exploring the molecular PES requires a capable MLP, which - in turn - requires representative training data.
In this work, every dataset is constructed post-AL, i.e., after we verified the model has become suitably accurate and samples the correct distribution of structural configurations. \\

\underline{Active learning:} in the AL workflow, exploration consists of 500 fs OpenMM walks using applied pressures randomly chosen between $-1.5$ and $1.5$ GPa. We restrict descriptor $\bm{\mathcal{F}}$ to the final hidden layer of every MLP, which has 8 or 16 dimensions (see Section SI.2) and directly precedes the atomic energy prediction.
Density models in feature space are parametrised using the Gaussian mixture implementation and expectation-maximisation algorithm of scikit-learn. \cite{ref_pedregosa_sklearn_2011} To design molecular clusters for different types of spatial disorder, we follow the approach illustrated in Section SI.6. \\

\underline{Error metrics:} as seen in Section \ref{sec_results_unit}, MLP force errors tend to be very localised around regions of disorder. Conventional metrics such as the mean absolute error (MAE) or root-mean-square error (RMSE) lack the sensitivity required to capture this local behaviour, while a maximal error is very susceptible to outliers. We propose a new metric:
\begin{equation}
\text{MAE}_P(X, k) = \text{MAE}(X') \text{ with } X' = \{x \in X | P_k (X) \le |x|\},
\end{equation}

in which $X$ is a multidimensional array of scalar error labels and $P_k (X)$ represents the k-th percentile of absolute values in $X$. The $\text{MAE}_P(X, k)$ is a thresholded MAE, computed over error magnitudes larger than $P_k (X)$. Setting a value of $k$ tunes the sensitivity to outlying values. In the limit, $k = 0$ reverts to the standard MAE and $k=100$ returns the maximal error. We choose $k=95$ and use $\text{MAE}_{P95}$ as the prime metric to discuss force accuracy in disordered structures. \\

Some MLPs make wildly inaccurate energy predictions for out-of-dataset systems. In these instances, the dominating source of error is usually a constant offset and the remaining variance  is very small. Because MAE or RMSE statistics fail to separate both error contributions,  we will report the mean and standard deviation of $\Delta E_i$:
\begin{align}
\Delta E_{\text{avg}} &= \frac{1}{M} \sum_i^M \Delta E_i , & \Delta E_{\text{std}} &= \sqrt{ \frac{1}{M} \sum_i^M (\Delta E_i - \Delta E_{\text{avg}})^2 }
\end{align}

where $\Delta E_i$ is the per-atom energy error for structure $i$ (out of $M$). For single-system datasets, $\Delta E_{\text{avg}}$ approximates the inherent shift between the MLP PES and \textit{ab initio} LOT. $\Delta E_{\text{std}}$ can be interpreted as an offset-corrected energy RMSE. The former metric can be ignored when comparing configurations of $S$; it is irrelevant for optimisations or MD sampling. On the contrary, $\Delta E_{\text{avg}}$ and $\Delta E_{\text{std}}$ are both important for accurate analysis of the relative stability of different systems. \\

\underline{Mechanical characterisation:} we investigate the mechanical behaviour of system $S$ by deriving static energy-vs-volume (EV) and dynamic pressure-vs-volume (PV) profiles to inspect the impact of defects on its properties.
EV curves are computed following the approach outlined in \cite{ref_vanpoucke_mechanical_2015}:
perform a series of fixed-volume structure optimisations for a grid of volume points - allowing cell shape and atomic positions to relax - and fit an appropriate equation-of-state (EOS). \cite{ref_birch_finite_1947,ref_vinet_compressibility_1987}
The bulk modulus of $S$, at zero kelvin, is found from the curvature of the resulting $E(V)$ relation.
More details can be found in Section SI.9.2. \\

PV curves are constructed at a finite temperature and explain the pressure response of $S$.
In the elastic strain regime, we perform NPT MD over a grid of pressures in OpenMM to find the equilibrium volume under applied pressure $\langle V (P_{\text{ext}}) \rangle$.
In unstable PV regions, we switch to the $(N, V, \bm{\sigma_a = 0}, T)$ ensemble implemented in YAFF \cite{ref_rogge_comparison_2015}, which constrains cell volume but allows its shape to vary freely, to recover the average internal pressure at a specified volume $\langle P_{\text{int}} (V) \rangle$.
Under equilibrium conditions, both ensembles should agree and, when combined, describe the complete PV behaviour.
From a PV curve of $S$, we can deduce its vacuum equilibrium volume, its bulk modulus and the maximal pressure it can withstand before collapsing.
Section SI.9.1 contains an in-depth explanation.

\section{Results} \label{sec_results}

Following a theoretical exposition in Section \ref{sec_methods}, we apply our AL workflow to several spatially disordered MOFs belonging to the UiO type series and answer the following questions:
(i) When do framework defects lead to large MLP inference inaccuracies?
(ii) Can we avoid out-of-dataset extrapolation with isolated chemical environments in molecular fragments?
(iii) Can cluster-based learning deliver transferable models for a family of disordered MOFs? \\

We select the prototypical zirconium MOF UiO-66(Zr) as the principal material from which defective structures will be derived.
It consists of Zr\textsubscript{6}O\textsubscript{4}(OH)\textsubscript{4} inorganic bricks connected with 12 1,4-benzenedicarboxylate (BDC) ligands, to create a network of coordination bonds adopting the fcu topology. \cite{ref_cavka_new_2008} UiO-66 is known for its tolerance to significant defect concentrations and is widely studied in both experimental and computational literature.
This provides ample reference data regarding its mechanical behaviour and the impact of various expressions of spatial disorder on framework properties. \\

From experiment, we identify three point defects that appear commonly in as-synthesized UiO-66, or that can be deliberately introduced with specialised synthesis protocols. \cite{ref_xiang_synthesis_2020,ref_meekel_correlated_2021,ref_cao_defect_2021}
Linker defects are missing organic ligands that create small vacancies in the regular topological lattice.
A metal substitution occurs when a chemically similar but different metal occupies an ionic atom site - conventionally hafnium or cerium for Zr-MOFs. \cite{ref_rogge_charting_2020}
Lastly, a node defect is a large framework void caused by the absence of a brick and all surrounding linkers.
Diffraction measurements show that such defects tend to appear in correlated nanodomains, forming local regions with reo topology. \cite{ref_cliffe_correlated_2014} \\

With these types of spatial disorder, we will construct disordered UiO-66 systems from the nanoscale (Section \ref{sec_results_unit}) to the mesoscale (Section \ref{sec_results_super}) and demonstrate how one can attain high-accuracy MLPs. In Section \ref{sec_results_props}, the superiour model will be used to describe the mechanical pressure response of a handful of UiO-66 variants, uncovering important defect-property relations.
In this work, we always restrict ourselves to periodic representations of bulk materials and do not consider any crystal surface phenomena.
First, however, we investigate the nature of MLP feature space (Section \ref{sec_results_feature_space}). \\

The following sections will compare multiple systems, datasets and MLPs. Section SI.10 provides an exhaustive overview for clarity.

\subsection{Chemical interpretation of MLP feature space} \label{sec_results_feature_space}

In Section \ref{sec_uncertainty}, we rely on a descriptor $\bm{\mathcal{F}}$ to deduce a metric of model uncertainty for sample $\epsilon$.
The underlying assumption is that MLP feature space inherently contains physical or chemical information about the neighbourhood of atoms.
Here, we will show that this embedded space is indeed informative for atomic environments. \\

We choose a conventional 456-atom unit cell, containing four bricks and 24 linkers, to represent the pristine UiO-66(Zr) framework and name it $S_{\text{pr}}$.
We generate a training ($D^T_{\text{pr}}$) and test ($D_{\text{pr}}$) dataset, containing 200 and 100 configurations of $S_{\text{pr}}$, respectively.
The superscript $T$ will consistently refer to a training set.
All structures are sampled uniformly in a volume range of $8500-9700$ $\text{\AA}^3$.
We train our first model on $D^T_{\text{pr}}$, label it \textbf{mlp}\textsubscript{pr}, and examine the structure of its (eight-dimensional) $\bm{\mathcal{F}}$-space. \\

\begin{figure}
\centering
\includegraphics[width=\linewidth]{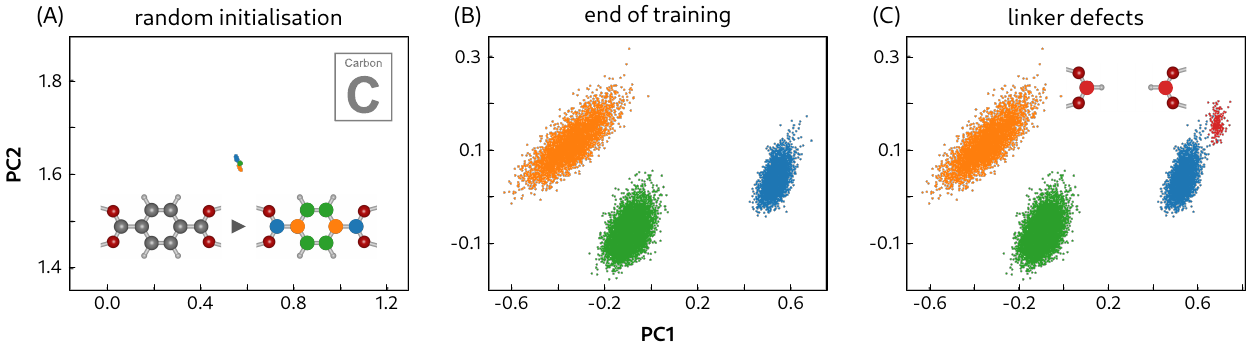}
\caption{Feature embeddings of C in UiO-66 ($D_{\text{pr}}$) for a randomly initialised model (A) and the fully trained \textbf{mlp}\textsubscript{pr} (B). In (C), $\bm{\mathcal{F}}$-vectors of linker defects are superimposed on panel (B). Datapoints are colour-coded according to their C-atom type.}
\label{fig_feature_space}
\end{figure}

We limit this discussion to $\epsilon$ and $\bm{\mathcal{F}}$ of carbon atoms in UiO-66.
First, we define a unique PCA reduction by extracting all C $\bm{\mathcal{F}}$-vectors of $D_{\text{pr}}(\epsilon)$ using \textbf{mlp}\textsubscript{pr}.
Every datapoint shown in Figure \ref{fig_feature_space} is projected on its two largest principal components.
Note that a 2D representation is less informative than the original 8D features, and is only performed for visualisation purposes. \\

Figure \ref{fig_feature_space}.A shows the $\bm{\mathcal{F}}$-embeddings of $D_{\text{pr}}(\epsilon)$, evaluated using a randomly initialised (i.e., untrained) checkpoint of \textbf{mlp}\textsubscript{pr}.
The scale of this plot matches Figure \ref{fig_feature_space}.B, which illustrates $D_{\text{pr}}(\epsilon)$ according to (the final checkpoint of) \textbf{mlp}\textsubscript{pr}.
Initially, the model projects all $\epsilon$ to a small region of $\bm{\mathcal{F}}$-space.
Throughout the training procedure, it learns to spread and separate $\epsilon$ to better reproduce the atomic interactions of $D^T_{\text{pr}}$.
Based on first-neighbour chemical intuition, we find 3 types of C in the BDC linker (see inset in Figure \ref{fig_feature_space}.A).
We colour-code every $\bm{\mathcal{F}}_i$ per carbon type of atom $i$ and observe that \textbf{mlp}\textsubscript{pr} groups $\epsilon$ in the same manner.
The division is never imposed on the model, hence it must have learned to encode this chemistry in its features. \\

In the next step, we introduce a defect in $S_{\text{pr}}$ by replacing a linker with two formate capping groups.
For a set of defective structures, \textbf{mlp}\textsubscript{pr} produces a fourth $\bm{\mathcal{F}}$-cloud (coloured red in Figure \ref{fig_feature_space}.C), distinct from existing ones, indicating a new type of carbon $\epsilon$.
The model has never encountered formate during training, but successfully distinguishes these $\epsilon$ from regular BDC carbon atoms, proving its feature space can recognise unknown $\epsilon$. \\

We stress that MLP $\bm{\mathcal{F}}$-space is highly nonlinear and model-specific, i.e., different models can give strongly divergent embeddings.
As such, attaching concrete chemical meaning to various regions of feature space is not really viable.
Nevertheless, qualitative characteristics - like the point cloud grouping in Figure \ref{fig_feature_space} - emerge naturally with model optimisation and conform to our intuition of different chemical environments.

\subsection{Disorder in UiO-66 unit cells} \label{sec_results_unit}

\begin{figure}
\centering
\includegraphics[width=\linewidth]{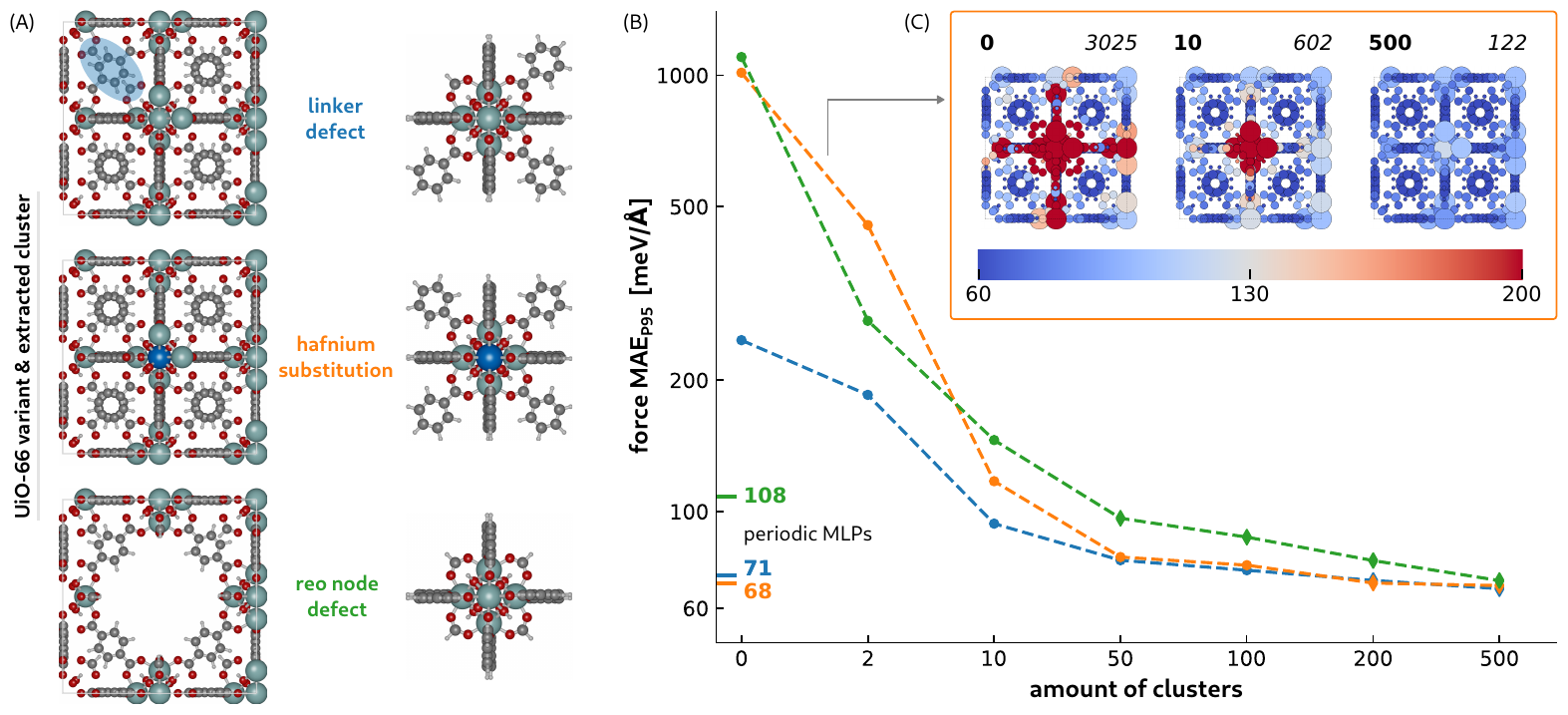}
\caption{Learning point defects in a UiO-66 unit cell.
    (A) An overview of $S_{\text{ld}}$, $S_{\text{hf}}$ and $S_{\text{reo}}$ and a matching molecular fragment to capture new $\epsilon$.
    (B) Learning curves showing force $\text{MAE}_{P95}$ versus $N$ for $D_{\text{ld}}$, $D_{\text{hf}}$ and $D_{\text{reo}}$ (see main text). When a curve undercuts its periodic counterpart, the marker changes from a dot to a diamond (see vertical axis).
    (C) A per-atom visualisation of force $\text{MAE}_{P95}$ for snapshots of the $D_{\text{hf}}$ learning curve.
    Above each cell, $N$ is indicated in bold, and the Hf atom force error is in cursive.}
\label{fig_defects}
\end{figure}

As a first case study, we investigate how introducing single point defects alters atomic interactions in UiO-66(Zr).
Finding a causal relation with the underlying change in $\epsilon$ can establish an informed pathway to construct representative datasets and train transferable MLPs. Starting from $S_{\text{pr}}$, we create three disordered systems by introducing a single linker defect (ld), a single hafnium substitution (hf) or a single node defect (reo).
Although literature offers different hypotheses regarding the termination of bricks in the absence of coordination-bonded ligands \cite{ref_cliffe_correlated_2014,ref_vandichel_active_2015}, we will consistently terminate linker defects with formate groups.
The resulting periodic structures are depicted in Figure \ref{fig_defects}.A and are referred to as $S_{\text{ld}}$, $S_{\text{hf}}$ and $S_{\text{reo}}$.
Table \ref{table_abbreviation} summarises some naming abbreviations commonly used in the following sections. \\

Analogous to $S_{\text{pr}}$, we generate periodic training and test datasets for every disordered UiO-66 variant ($D^T_{\text{ld}}$ and $D_{\text{ld}}$, $D^T_{\text{hf}}$ and $D_{\text{hf}}$, $D^T_{\text{reo}}$ and $D_{\text{reo}}$). Our baseline model, \textbf{mlp}\textsubscript{pr}, performs excellently for $D_{\text{pr}}$ with $\Delta E_{\text{std}} < 0.4$ meV/atom and force RMSE $< 26$ meV/\AA.
However, it fails to correctly reproduce molecular interactions in the neighbourhood of spatial defects for any other test set (see Figure \ref{fig_defects}.C and Figure SI.10 or Table SI.4 and Table SI.5).
Notably, model errors in framework regions free from disorder remain very low.
The localised nature of these extreme force errors points to the existence of new $\epsilon \notin D^T_{\text{pr}}(\epsilon)$.
In UiO-66, the qualitative `area of effect' (AOE) of a linker defect, i.e., how many atoms feel the missing linker, is relatively small, followed by a hafnium substitution and a node defect, affecting almost the entire unit cell. \\

To address deficiencies in $D^T_{\text{pr}}(\epsilon)$ and improve the transferability of \textbf{mlp}\textsubscript{pr}, we employ the cluster-based learning methodology.
The design principles of Section \ref{sec_extraction} readily identify three suitable molecular fragments that isolate and extract new $\epsilon$ from $S_{\text{ld}}$, $S_{\text{hf}}$ and $S_{\text{reo}}$ (see Figure \ref{fig_defects}.A).
We initiate an AL campaign with \textbf{mlp}\textsubscript{pr} and the three disordered systems as learning targets.
Every cycle, 50 walks are performed per system and 150 different cluster conformations are gathered to retrain the MLP. After several iterations, outlying force errors largely vanish and the new model successfully learns to describe simple defects.
As points of reference, we also train MLPs for each of the disordered unit cells: \textbf{mlp}\textsubscript{ld} on $D^T_{\text{ld}}$, \textbf{mlp}\textsubscript{hf} on $D^T_{\text{hf}}$ and \textbf{mlp}\textsubscript{reo} on $D^T_{\text{reo}}$.
These `periodic' models - trained solely on periodic structures - will be compared with `cluster' MLPs, which include non-periodic training data as well. \\

We construct a learning curve for $D_{\text{ld}}$, $D_{\text{hf}}$ and $D_{\text{reo}}$ to systematically study model accuracy with varying amounts of cluster training data in Figure \ref{fig_defects}.B.
First, all extracted fragments from the AL campaign are amassed into a database, from which we sample random subsets of size $N$ (between 2-500) for ld, hf and reo defects separately.
Every cluster subset is combined with $D^T_{\text{pr}}$ to retrain \textbf{mlp}\textsubscript{pr}, resulting in a single datapoint.
Figure \ref{fig_defects}.B shows the full learning curves and plots force $\text{MAE}_{P95}$ (see Section \ref{sec_comp_det}) on a logarithmic scale versus $N$ - the number of clusters added to $D^T_{\text{pr}}$.
All curves start from \textbf{mlp}\textsubscript{pr} ($N = 0$) and gradually incorporate more clusters of the corresponding defect type.
Improvements in model accuracy are a direct consequence of the newly included $\epsilon$.
The $\text{MAE}_{P95}$ metrics of periodic models (\textbf{mlp}\textsubscript{ld}, \textbf{mlp}\textsubscript{hf} and \textbf{mlp}\textsubscript{reo}) are indicated on the vertical axis of Figure \ref{fig_defects}.B with coloured dashes. \\

We observe that \textbf{mlp}\textsubscript{reo} performs strikingly worse than \textbf{mlp}\textsubscript{ld} and \textbf{mlp}\textsubscript{hf} on their respective dataset.
Describing the large void created by a reo defect correctly might be intrinsically more difficult. \cite{ref_behler_machine_2021}
Node defects have a large AOE: many $\epsilon$ contribute to the measured error. $S_{\text{reo}}$ also contains significantly fewer atoms, making $D^T_{\text{reo}}$ the smallest dataset.
This is counterbalanced by the fact that most $\epsilon$ in $S_{\text{reo}}$ contain useful information about the missing node (compare with the AOE of a linker defect in $S_{\text{ld}}$).
Increasing network complexity and $r_{\text{max}}$ results in a more performant model for the same $D^T_{\text{reo}}$.
However, we use model error as a proxy to compare and find flaws in $D^T(\epsilon)$.
Therefore, we must eliminate extraneous variables and keep the NequIP hyperparameter configuration fixed for all models. \\

In Figure \ref{fig_defects}.B, \textbf{mlp}\textsubscript{pr} is markedly more inaccurate for $D_{\text{hf}}$ and $D_{\text{reo}}$ than for $D_{\text{ld}}$.
Large errors for $D_{\text{hf}}$ are unsurprising; the model has never encountered the element Hf and will guess randomly in its vicinity.
The difference between $D_{\text{reo}}$ and $D_{\text{ld}}$ can be understood from their relative AOE, i.e., missing linkers are much more local.
Every learning curve quickly surpasses its matching periodic MLP - after 50 clusters for $S_{\text{reo}}$ and after 200 clusters for $S_{\text{ld}}$ and $S_{\text{hf}}$.
This proves that the fragments of Figure \ref{fig_defects}.A properly capture relevant $\epsilon$ from their parent systems.
We observe the biggest improvements in accuracy for low $N$ ($< 50$). Including additional clusters leads to diminishing returns, a pattern that is expected in machine learning. \cite{ref_frey_neural_2023} \\

Figure \ref{fig_defects}.C provides a per-atom visualisation of force $\text{MAE}_{P95}$ for $D_{\text{hf}}$ along three points of its learning curve, $N \in \{0, 10, 500\}$ (indicated in bold).
At $N=0$, large force errors coalesce in a sizeable sphere centered on the Hf substitution.
As more training clusters are added, the sphere steadily shrinks in radius and error magnitudes decrease.
$\text{MAE}_{P95}$ values for the single Hf atom (reported in cursive) drop from a massive 3025 meV/\AA\ to just 122 meV/\AA.
Still, the average $\text{MAE}_{P95}$ for Zr atoms is only 97 meV/\AA, owing to the Hf/Zr ratio in the final dataset.
Similar observations hold for linker and node defects (see Figure SI.10 and Table SI.10 for average error values). \\

From these results, we conclude that adapting an existing MLP, \textbf{mlp}\textsubscript{pr}, with a modest amount of clusters is a viable alternative to training new periodic models from scratch.
Moreover, cluster-based learning can lower the \textit{ab initio} computational cost of dataset generation significantly - roughly by a factor of five in this case - without forfeiting model accuracy.
This advantage will amplify as periodic cells grow larger or the level of theory becomes more expensive. \\

Up to now, we have only evaluated models on their specific defect type (ld, hf and reo).
In Section SI.8, we cross-validate MLPs and test datasets to uncover relations between different kinds of spatial disorder.
We find that force errors are more sensitive to MLP extrapolation than energy errors, and should be preferred when trying to identify missing interactions.
In inference, `cluster' models for $S_{\text{ld}}$ and $S_{\text{reo}}$ perform very similarly, indicating a strong correspondence in $\epsilon$ for linker and reo defects.
The most transferable model, named \textbf{mlp}$^c_{\text{mix}}$, is obtained by combining all three cluster types (see Table SI.9).
A superscript $c$ indicates the MLP is trained using clusters, alongside the basic $D^T_{\text{pr}}$ dataset.
Energy errors generally decompose into small $\Delta E_{\text{std}}$ and enormous $\Delta E_{\text{avg}}$ values.
For most models, the absolute energy scale is clearly wrong, although relative energy differences are captured adequately.
Only \textbf{mlp}$^c_{\text{mix}}$ manages to accurately predict absolute energies for all test sets, owing to the compositional variety of its training data.

\subsection{Disorder in a UiO-66 supercell} \label{sec_results_super}

\begin{figure}
\centering
\includegraphics[width=0.8\linewidth]{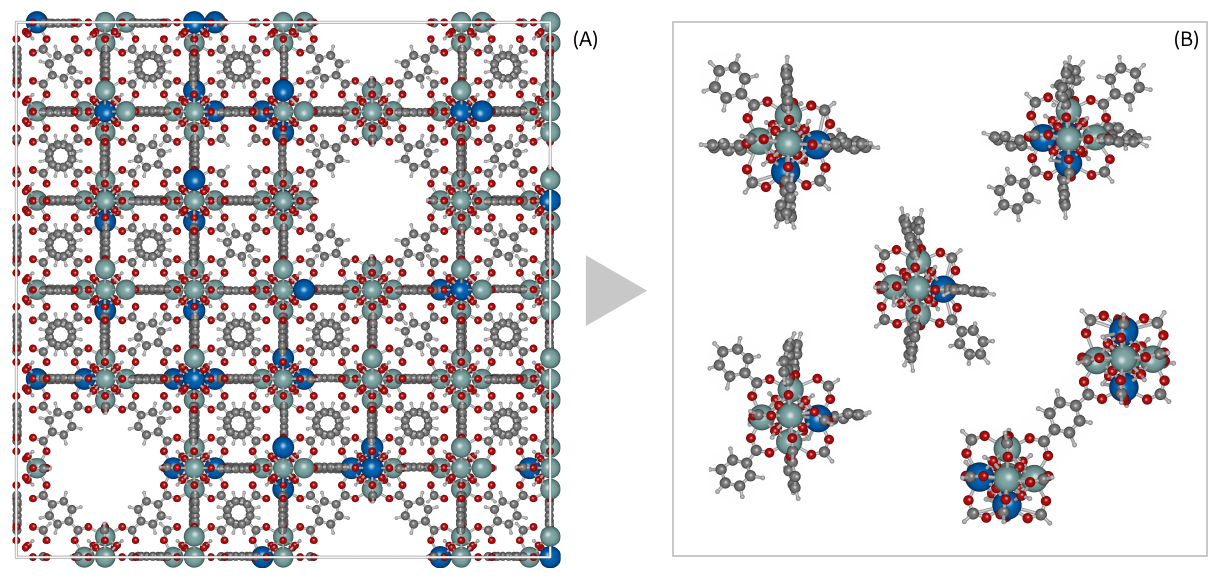}
\caption{Spatial disorder in UiO-66 at the mesoscale. (A) An example 3x3x1 supercell similar to $S_{\text{sup}}$. (B) A handful of clusters extracted from $S_{\text{sup}}$.}
\label{fig_supercell}
\end{figure}

In Section \ref{sec_results_unit}, we investigated spatial disorder as individual point defects in UiO-66 unit cells.
So far, periodic correlations heavily limited the available configurational freedom of atoms and disorder, and hence the explorable $\epsilon$-space.
To more closely approximate realistic frameworks, an ensemble of missing linkers, metal substitutions and node defects should be considered.
We generate a highly disordered UiO-66 system, starting from a pristine 4x4x4 supercell (29184 atoms), by randomly removing 20\% of linkers, replacing 20\% of Zr atoms with Hf and creating node defects from 10\% of bricks and surrounding ligands.
Dangling bonds are saturated by hydrogen to form formate groups (analogous to Section \ref{sec_results_unit}) and unconnected building blocks are discarded from the remaining framework.
The resulting structure contains 22052 atoms and will be labelled $S_{\text{sup}}$.
Since the distribution of disorder is completely random and the concentration of defects is very substantial, we expect to find a wide variety of previously unexplored environments, such as fully mixed Zr-Hf bricks and mesoporous channels caused by adjacent reo cavities.
A 3x3x1 example cell, identically constructed to $S_{\text{sup}}$, is shown in Figure \ref{fig_supercell}.A. \\

We commence an AL campaign to assemble representative data for $S_{\text{sup}}$.
The initial training set remains $D^T_{\text{pr}}$ - i.e., the pristine 456-atom unit cell - to ensure consistency with previous models.
However, we expand the MLP architecture of \textbf{mlp}\textsubscript{pr} towards more internal parameters and an increased $r_{\text{max}}$ (see Section SI.2), anticipating a vastly enlarged and more complex $S_{\text{sup}}(\epsilon)|_{\text{ENS}}$.
In every AL iteration, we perform 2 MD simulations and collect roughly 150 fragments, extracted following the design rules from Section SI.6.
Figure \ref{fig_supercell}.B depicts a handful of examples.
After a dozen AL rounds, we select 1500 clusters (like the training set of \textbf{mlp}$^c_{\text{mix}}$), retrain the AL model and name it \textbf{mlp}$^c_{\text{sup}}$. \\

\begin{table}[]
\includegraphics[height=5cm]{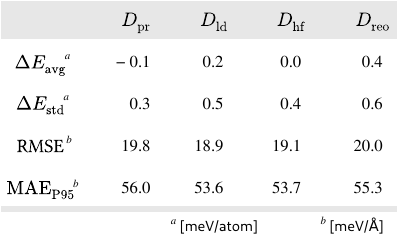}
\caption{Validation metrics of \textbf{mlp}$^c_{\text{sup}}$ for every test dataset from Section \ref{sec_results_unit}.}
\label{table_val_super_a}
\end{table}

\begin{table}[]
\includegraphics[height=5cm]{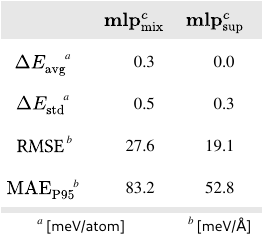}
\caption{Validation metrics of \textbf{mlp}$^c_{\text{mix}}$ and \textbf{mlp}$^c_{\text{sup}}$ for $D_\text{cl}$.}
\label{table_val_super_b}
\end{table}

Because \textit{ab initio} calculations of $S_{\text{sup}}$ are computationally infeasible, we resort to our unit cell test sets to evaluate model performance.
Table \ref{table_val_super_a} and Table SI.11 show virtually identical energy errors between \textbf{mlp}$^c_{\text{mix}}$ and \textbf{mlp}$^c_{\text{sup}}$; we appear to have reached the accuracy limit for energy predictions with our DFT/MLP configuration.
In contrast, force RMSE and $\text{MAE}_{P95}$ improve by roughly 13-24\% depending on the dataset and metric.
By construction, both models differ in hyperparameters and training data.
To isolate the effects of each difference, we train a final MLP with the architecture of one model (\textbf{mlp}$^c_{\text{mix}}$) and the dataset of the other (\textbf{mlp}$^c_{\text{sup}}$).
It still surpasses \textbf{mlp}$^c_{\text{mix}}$ in force inference, although relative improvement shrinks to 0-9\% (see \textbf{mlp}$^c_{\text{sup*}}$ in Table SI.11).
We conclude that the superiour accuracy of \textbf{mlp}$^c_{\text{sup}}$ is caused in part by a more expressive dataset, but mostly by a larger network size.
Our best model for $D_{\text{ld}}$, $D_{\text{hf}}$ and $D_{\text{reo}}$ is derived from clusters that were not extracted from configurations of $S_{\text{ld}}$, $S_{\text{hf}}$ or $S_{\text{reo}}$.
The most representative dataset contains the greatest diversity in $\epsilon$, regardless of the fragments' parent system (compare Figure \ref{fig_defects}.A and Figure \ref{fig_supercell}.B). \\

Validation metrics on simple point defects might not generalise to arbitrarily disordered frameworks.
As a substitute for $S_{\text{sup}}$, we construct a cluster test set $D_\text{cl}$ containing 500 fragments, newly sampled from MD simulations.
In Table \ref{table_val_super_b}, \textbf{mlp}$^c_{\text{sup}}$ outperforms \textbf{mlp}$^c_{\text{mix}}$ in all error statistics, and more convincingly than in Table \ref{table_val_super_a}.
Most surprising, however, is the robustness of \textbf{mlp}$^c_{\text{mix}}$ for (out-of-dataset) clusters.
This model never learned interactions between defects during training - e.g., multiple hf and ld defects in a brick - yet remains impressively accurate on $D_\text{cl}$.
A posteriori, we discover that \textbf{mlp}$^c_{\text{mix}}$ can describe the PES of $S_{\text{sup}}$ decently well, or equivalently, that all essential $\epsilon$ in $S_{\text{sup}}$ could be learned with clusters from $S_{\text{ld}}$, $S_{\text{hf}}$ and $S_{\text{reo}}$.
Based on isolated Hf substitutions, it inferred that Zr and Hf serve a similar role in UiO-66.
Nevertheless, force RMSE values for Zr and Hf atoms are 44.5 and 68.7 meV/\AA\ for \textbf{mlp}$^c_{\text{mix}}$, and 28.9 and 29.3 meV/\AA\ for \textbf{mlp}$^c_{\text{sup}}$, indicating that (interactions of) metal ions limit overall accuracy in \textbf{mlp}$^c_{\text{mix}}$.
Note that $D_\text{cl}$ error metrics are only indicative of performance for the original periodic system $S_{\text{sup}}$.
Molecular clusters should capture interactions from their parent, but this is difficult to verify at the mesoscale. \\

Combining conclusions from Section \ref{sec_results_unit} and Section \ref{sec_results_super}, we summarise: for an MLP and training set $D^T$,
(i) defects or interactions wholly absent from $D^T$ cause huge local force errors (Figure \ref{fig_defects}.B),
(ii) combinations of defects are likely not problematic if every type of disorder is contained in $D^T$ separately (Table \ref{table_val_super_b}), and
(iii) the most accurate and transferable model is trained from the most extensive $D^T(\epsilon)$. Constructing such $D^T(\epsilon)$ can be challenging and requires trial-and-error or hands-on experience with the material of interest. Our cluster-based learning algorithm abstracts the required know-how and delivers representative datasets for arbitrarily disordered systems.

\subsection{Mechanical properties of disordered UiO-66 species} \label{sec_results_props}

Creating performant atomic potentials is only the first step in uncovering structure-property relations of materials.
In this section, we will employ \textbf{mlp}$^c_{\text{sup}}$ to investigate the mechanical behaviour of several spatially disordered UiO-66 species, which also serves as a first order validation to show reliable MLP predictions of derived properties.
Starting from the pristine UiO-66(Zr) unit cell, $S_{\text{pr}}$, we systematically incorporate higher concentrations of disorder. $S_{\text{ld}}$ is created by introducing a single linker defect (see Section \ref{sec_results_unit}). We form cells with an average brick coordination number of 11 by removing two ligands.
Rogge et al. showed that the crystal symmetry of UiO-66 allows for seven physically distinct configurations of a double linker defect (see Figure SI.8); each with its own set of properties. \cite{ref_rogge_thermodynamic_2016}
We will not fixate on an in-depth comparison between these variants, hence, refer to them collectively as `ld-2' systems ($S_\text{ld-2}^\text{1-7}$).
Removing a third ligand would generate a combinatorially exploding number of new structures.
Instead, we will focus on three framework topologies recently observed in transmission electron microscopy experiments (see Figure SI.9).
\cite{ref_liu_imaging_2019}
A bcu network, $S_\text{bcu}$, is obtained by removing all linkers in planes perpendicular to a chosen cell axis (X, Y, or Z), leaving four 8-connected bricks and 16 linkers.
We make $S_{\text{reo}}$, with three 8-connected bricks and 12 linkers, using a reo-type node defect.
Superimposing the defects of $S_\text{bcu}$ and $S_{\text{reo}}$ results in the scu topology, $S_\text{scu}$, characterized by one 8-connected and two 4-connected bricks held together with 8 linkers.
Finally, we include a pristine UiO-66(Hf) unit cell as a reference point for hafnium-substituted materials ($S_\text{pr}^\text{hf}$). \\

\begin{figure}
\centering
\includegraphics[width=0.8\linewidth]{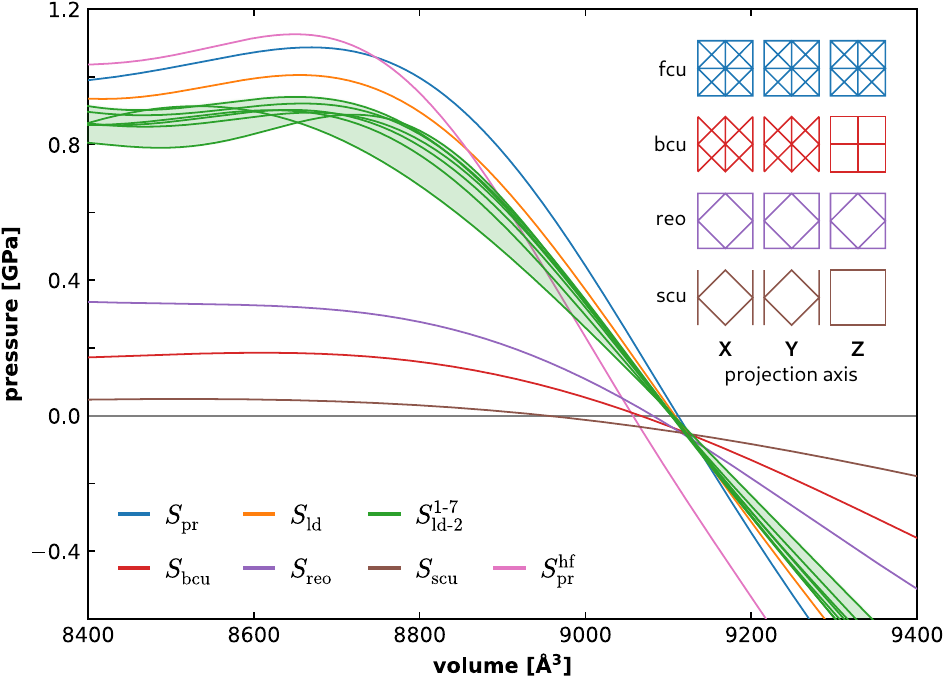}
\caption{Pressure-vs-volume curves computed for a diverse set of UiO-66-derived systems (see main text). Frameworks with double linker defects ($S_\text{ld-2}^\text{1-7}$) are merged in green. The inset shows structural representations of $S_{\text{pr}}$, $S_\text{bcu}$, $S_{\text{reo}}$ and $S_\text{scu}$, (see colours) projected along principal cell axes to highlight differences in topology.}
\label{fig_pv}
\end{figure}

\begin{table}[]
\centering
\includegraphics[width=0.32\linewidth]{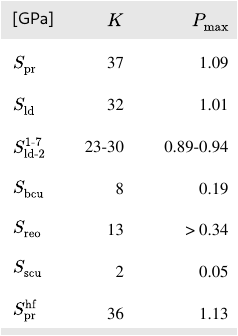}
\caption{Bulk moduli and loss-of-crystallinity pressures derived from Figure \ref{fig_pv}. For $S_\text{ld-2}^\text{1-7}$, the reported values represent an aggregated range.}
\label{table_pv}
\end{table}

Figure \ref{fig_pv} shows the pressure-vs-volume behaviour at room temperature (300 K) of all aforementioned systems, computed with \textbf{mlp}$^c_{\text{sup}}$ following the procedure in Section \ref{sec_comp_det}. Table \ref{table_pv} reports the bulk modulus $K$ at equilibrium volume $V_0$ and PV maximum $P_\text{max}$ for every curve. $P_\text{max}$ is the maximal pressure a system can withstand before collapsing into an unstable PV branch.
It coincides with a significant drop in internal symmetry for MOFs and is sometimes called the loss-of-crystallinity pressure. \cite{ref_rogge_thermodynamic_2016}
At lower cell volumes, we expect to recover another stable branch corresponding with the compression of an amorphous framework.
Note that the interpretation of $K$ under anisotropic strains is not straightforward.
Nevertheless, it forms a starting point to compare the mechanical properties of our systems with earlier computational and experimental predictions. \\

For $S_{\text{pr}}$, we find a bulk modulus of 37 GPA, which is in good agreement with static \textit{ab initio} predictions of 41-42 GPa \cite{ref_wu_exceptional_2013,ref_redfern_isolating_2020},
and in even better agreement with an experimental study that found 37.9 GPa using in situ synchrotron X-ray powder diffraction. \cite{ref_redfern_porosity_2019} In the same experiment, $V_0$ was measured at 9009 \AA\textsuperscript{3}, which we overshoot by roughly 100 \AA\textsuperscript{3} ($\pm1\%$) - a known consequence of the PBE functional approximation. \cite{ref_lejaeghere_error_2014}
Our simulations suggest that pristine UiO-66 will collapse under hydrostatic pressures above 1.1 GPa. From literature, one expects mechanical resilience to decrease when incorporating (linker) defects into MOFs. \cite{ref_dissegna_tuning_2018,ref_thornton_defects_2016,ref_vervoorts_structural_2021}
However, interpolating consistent quantitative results rather than qualitative trends across various sources is difficult.
Empirical amorphisation seems particularly troublesome in this regard, as a measurable analogue of $P_\text{max}$ is hard to define.
Figure \ref{fig_pv} predicts a modest drop in both $K$ and $P_\text{max}$ for $S_{\text{ld}}$.
We find bulk moduli between 23-30 GPa and loss-of-crystallinity pressures around 0.89-0.94 GPa for the various $S_\text{ld-2}^\text{1-7}$ systems, emphasising that material characteristics are governed by the concentration as well as the distribution of spatial disorder.
In Section SI.9.3, we compare our findings with earlier work by Rogge et al. \cite{ref_rogge_thermodynamic_2016}, which employs system-specific forcefields parametrised through QuickFF \cite{ref_vanduyfhuys_extension_2018}, for the systems discussed so far.
We consistently predict larger bulk moduli and smaller loss-of-crystallinity pressures, but recover a robust linear relation for $K$ between both LOTs. \\

Alterations in framework topology induce drastic changes in mechanical properties. Figure \ref{fig_pv} shows a clear reduction in $V_0$ and $K$ following the order $S_\text{pr} \text{(fcu)} \rightarrow S_\text{reo} \rightarrow S_\text{bcu} \rightarrow S_\text{scu}$, which seems surprising at first, given that  $S_\text{bcu}$ retains more building blocks of the original fcu cell than $S_{\text{reo}}$. To explain this behaviour, the inset in Figure \ref{fig_pv} illustrates a schematic depiction of each topology by projecting its nodes and ligands along every cell axis. These structural representations show that the asymmetric removal of ligands creates a weak crystal axis in $S_\text{bcu}$ and $S_\text{scu}$, i.e., they will compress more easily in the XY-plane and elongate in the Z-direction under hydrostatic pressure. On the contrary, $S_{\text{reo}}$ is symmetric in all three major axes, meaning it has no preferred direction of strain. In terms of $P_\text{max}$, both $S_\text{bcu}$ and $S_\text{scu}$ show a slight maximum in the considered volume range, whereas $S_{\text{reo}}$ does not; its PV curve keeps rising steadily as the cell shrinks to volumes where MLP accuracy can no longer be assumed. Moreover, it exhibits (at least) two linear stable branches separated by a transitionary region between 8700-9000 \AA\textsuperscript{3}, potentially indicating the existence of multiple stable phases. At present, we can only estimate a lower bound on $P_\text{max}$ for $S_{\text{reo}}$ ($> 340$ MPa). In Section SI.9.4, we construct EV profiles for all topologies to uncover major structural changes that occur under compression. Our analysis indicates that cells of $S_{\text{pr}}$ and $S_{\text{reo}}$ compress through a collective rotation of building blocks and buckling of ligands, whereas $S_\text{bcu}$ and $S_\text{scu}$ undergo a limited reorientation of coordination bonds as a shearing strain. These distinct deformation mechanisms can explain the differences in mechanical behaviour of Figure \ref{fig_pv}.  \\

We conclude with UiO-66(Hf), $S_\text{pr}^\text{hf}$, for which DFT calculations predict a bulk modulus of 39.5 GPa and experimental measurements find $V_0$ and $K$ estimates of 8906 \AA\textsuperscript{3} and 37 GPa, respectively. \cite{ref_wu_exceptional_2013,ref_redfern_isolating_2020}
Both values are marginally smaller than those for the Zr framework; a trend our simulations reproduce. Additionally, $P_\text{max}$ is only slightly larger. We expect the mechanical properties of Zr-Hf UiO-66 mixtures to not deviate strongly from those observed for single-metal frameworks, which was already concluded in a recent forcefield study. \cite{ref_rogge_charting_2020} \\

\begin{figure}
    \centering
    \includegraphics[width=0.7\linewidth]{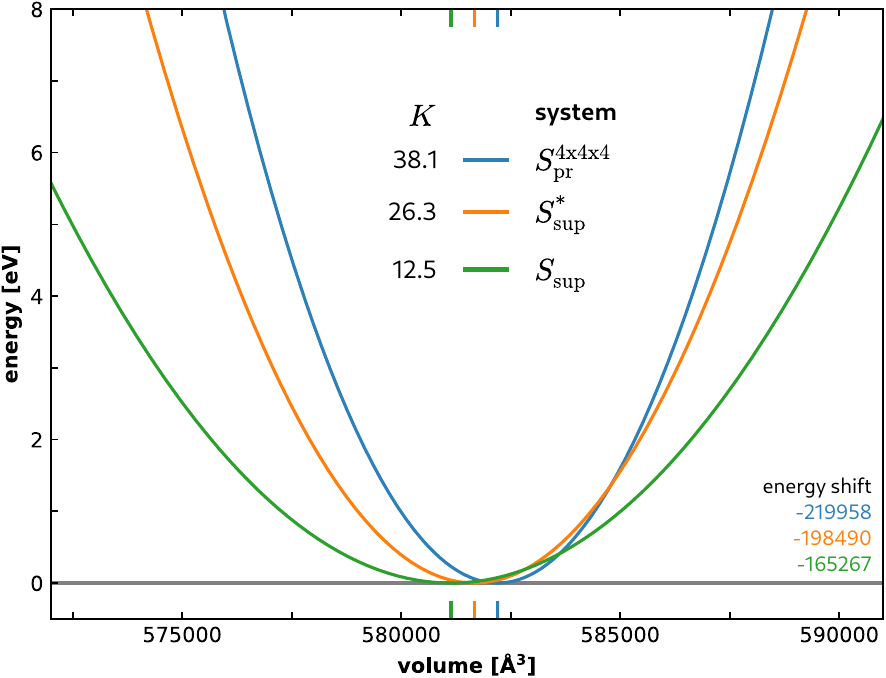}
    \caption{Energy-vs-volume curves for 4x4x4 UiO-66 supercells with varying degrees of spatial disorder (see main text), along with the corresponding energy baseline shifts, optimal cell volumes and  bulk moduli (in GPa).}
    \label{fig_eos_super}
\end{figure}

\underline{Mechanical behaviour of supercells:} in the discussion above, we restricted MOF systems to unit cell dimensions. However, we ultimately aim to describe realistic and disordered frameworks at the mesoscale. Here, we consider three 4x4x4 supercells of UiO-66 containing over 20k atoms, distinguished by their concentration of spatial disorder: an upscaled version of $S_{\text{pr}}$ ($S_{\text{pr}}^\text{4x4x4}$), $S_{\text{sup}}$, and a third system with roughly half the defects of $S_{\text{sup}}$ - named $S^*_{\text{sup}}$. Disorder is introduced at random. We adopt a simple characterisation scheme ($c_\text{hf}$ - $c_\text{brick}$ - $c_\text{linker}$), where $c_\text{hf}$ denotes the fraction of metal sites occupied by Hf atoms, and $c_\text{brick}$ and $c_\text{linker}$ represent the ratio of missing bricks and linkers compared to a defect-free system. With this convention, $S_{\text{pr}}^\text{4x4x4}$, $S^*_{\text{sup}}$ and $S_{\text{sup}}$ correspond with (0 - 0 - 0), (0.10 - 0.04 - 0.17) and (0.21 - 0.13 - 0.40) supercells, respectively. \\

Because simulating PV profiles becomes quite costly at these length scales, Figure \ref{fig_eos_super} shows EV curves for every system, evaluated with \textbf{mlp}$^c_{\text{sup}}$ and baseline shifted to remove energy offsets (see Section \ref{sec_comp_det}). As the degree of disorder increases, average brick coordination numbers lower from 12 to 10.4 and 8.3. Hand-in-hand, the associated bulk moduli drop threefold from 38.1 to 12.5 GPa, and we notice a limited amount of cell contraction ($< 0.2\%$), aligning with our earlier findings in unit cells (see Figure \ref{fig_pv}). For reference, EV profiles for $S_{\text{pr}}$, $S_{\text{reo}}$, $S_\text{bcu}$ and $S_\text{scu}$ find $K$ values of 38.1, 17.7, 10.7 and 4.9 GPa. We observe that $S_{\text{reo}}$, a (0 - 0.25 - 0.50) unit cell, has more node defects and a lower coordination number than any supercell considered, yet manages a remarkable resistance to compression. This impressive stability has been attributed to the correlated nature of defects in the reo topology. \cite{ref_thornton_defects_2016} \\

Note that our characterisation says nothing about the distribution of spatial disorder. For statistically representative EV profiles, we would need to average over an appropriate ensemble of disordered MOFs with a fixed defect concentration ($c_\text{hf}$ - $c_\text{brick}$ - $c_\text{linker}$). While this is overtly out-of-scope in a proof-of-concept work, we have shown that cluster-based learning provides the toolbox necessary to tackle such investigations.

\section{Discussion}

In this closing section, we reflect on the advantages and shortcomings of our methodology and discuss potential extensions and future research avenues. \\

Cluster-based learning enables MLP training for molecular systems at any length scale.
It is almost fully automated, can be seamlessly integrated into modular AL workflows and delivers transferable models.
Discrepancies between MLPs will inevitably propagate into differences of (mechanical) behaviour between systems.
The ability to describe all systems of interest with a single PES eliminates this source of inconsistency, allowing an apples-to-apples comparison of properties.
Currently, defining appropriate clusters requires a fair amount of trial and error (see Section SI.6).
In time, a set of base rules will be established that provides a workable initial guess for fragments, to be finetuned for specific use cases.
Our implementation deconstructs frameworks like the UiO-66 series into discrete building blocks; but should be adapted for more complex topologies, such as winerack MOFs or even other classes of materials.
Concerning uncertainty quantification, we will experiment with new combinations of feature descriptors $\bm{\mathcal{F}}$ and density models (GMM, as of now) to strengthen the relation between $\epsilon$-likelihood and MLP inference error. \\

To explore the behaviour of realistic frameworks with spatial disorder, large molecular structures with a suitable concentration and composition of defects are needed.
Experimentally, one cannot always probe the distribution of disorder, and we commonly assume that point defects are scattered homogeneously throughout the material.
Many exceptions exist, however, e.g., correlated reo-defect nanodomains in UiO-66. \cite{ref_cheetham_defects_2016,ref_cliffe_correlated_2014,ref_meekel_correlated_2021}
In Section \ref{sec_results_super} and Section \ref{sec_results_props}, we built test systems by randomly introducing point defects.
While this naive approach suffices to collect a large variety of $\epsilon$ for model training, it will not generate representative structures resembling synthesised crystallites.
As a first improvement, we could use Monte Carlo methods or the quasi-chemical approximation to create defective systems based on energetic and entropic grounds. \cite{ref_liu_monte_2021,ref_brivio_thermodynamic_2016,ref_sher_quasichemical_1987} \\

Cluster-based learning is most powerful when exorbitant computational costs prohibit \textit{ab initio} evaluations for a chosen system.
Even high-performance computing infrastructures struggle with the computational requirements posed by post-Hartree-Fock methods or DFT functionals higher up on Jacob's ladder for all but the smallest systems.
When studying MOFs, we usually resort to the GGA functionals.
These approximations cannot describe London dispersion forces and only crudely capture electron correlations, leading to significant underbinding and deviations from real-world properties. \cite{ref_lejaeghere_error_2014}
With clusters as training data, we can afford \textit{ab initio} calculations at otherwise inaccessible LOTs, effectively bypassing the quantum scaling limit.
In particular, $\Delta$-learning workflows, in which MLPs are trained to predict a low-cost LOT and a higher order correction (e.g., PBE to MP2), could benefit from this approach and potentially reach chemical accuracy for MOFs. \cite{ref_dral_hierarchical_2020}
A second accessible application is the study of populated frameworks.
Molecular fragments can isolate guest molecules and their immediate surroundings from the bulk MOF, enabling MLPs to learn and describe diffusion or adsorption processes.

\section{Conclusion}

This work explores the fundamental relations between mechanical properties and spatial disorder for a series of UiO-66(Zr)-derived frameworks.
We introduce the cluster-based learning methodology to develop robust MLPs at extended length scales.
It identifies unknown chemical environments in sample structures through MLP feature space.
These are extracted from the molecular bulk as compact fragments to extend the chemical space of atomic interactions covered by the model's training data.  \\

We use this method to learn various point defects in small unit cells.
Our investigation shows how to predict MLP extrapolation errors, how different types of spatial disorder are related and how to construct representative datasets that outperform conventionally trained models in accuracy and cost-effectiveness.
We employ our procedure to successfully train a performant model from a strongly disordered 4x4x4 supercell containing over twenty thousand atoms.
The major takeaway is that a greater variety of chemical environments in the training set delivers more accurate and transferable MLPs. \\

Using our leading model, we probe the pressure-versus-volume behaviour of pristine UiO-66, disordered cells with up to two linker defects and three experimentally observed framework topologies.
Finally, we extend our analysis to disordered supercells at the mesoscale, examining energy-versus-volume characteristics.
These simulations highlight the impact of the concentration, composition and correlated nature of spatial disorder on framework properties. \\

We have shown that cluster-based learning enables the development of highly authentic MLPs by evaluating and learning atomic interactions in small clusters.
Afterwards, these models can be applied on larger disordered systems, unlocking the study of MOFs and spatial disorder at unprecedented length scales, potentially including external surfaces.

\begin{acknowledgement}

    V.V.S. acknowledges funding from the Research Board of Ghent University (BOF).
    P.D. and S.V. wish to thank the Fund for Scientific Research-Flanders (FWO) for aspirant doctoral fellowships (grant nos. 11O2125N and 11H6821N).
    The computational resources and services used in this work were provided by the VSC (Flemish Supercomputer Center), funded by the Research Foundation - Flanders (FWO) and the Flemish Government department EWI.

\end{acknowledgement}

\begin{suppinfo}

    The Supporting Information is divided into several sections:
    \textit{Ab initio} calculations in GPAW,
    NequIP architecture and training setup,
    Molecular dynamics,
    Active learning,
    Cluster extraction,
    Example force matching,
    MLP uncertainty and force errors,
    MLP cross-validation,
    Mechanical characterisation,
    Overview of systems, datasets and MLPs,
    MLP accuracy and test metrics. \\

    Datasets, trained models and further data are made available through a Zenodo repository at DOI: \url{10.5281/zenodo.14846185}

\end{suppinfo}

\bibliography{refs.bib}

\end{document}


\begin{center}
    \renewcommand{\baselinestretch}{1.5}
    \huge
    Supporting information for \\
    \LARGE
    \textbf{Cluster-based machine learning potentials to describe disordered metal-organic frameworks up to the mesoscale} \\
    \vspace{20pt}
    \large
    Pieter Dobbelaere, Sander Vandenhaute and Veronique Van Speybroeck* \\
    \vspace{10pt}
    \textit{Center for Molecular Modeling (CMM), Ghent University, \\ Technologiepark 46, 9052 Zwijnaarde, Belgium} \\
    \vspace{10pt}
    E-mail: veronique.vanspeybroeck@ugent.be
\end{center}

\newpage
\tableofcontents

\addtocontents{toc}{\setcounter{tocdepth}{0}}

\setcounter{page}{3}
\addtocontents{toc}{\setcounter{tocdepth}{3}}

\setcounter{equation}{0}
\setcounter{section}{0}
\setcounter{table}{0}
\setcounter{figure}{0}

\section{\textit{Ab initio} caclulations in GPAW} \label{si_gpaw}

Structure evaluations at the DFT LOT are performed using the GPAW engine (version 22.8.0) \cite{ref_mortensen_2005}, which was chosen because our workflow involves both finite clusters and periodic structures. GPAW contains a finite grid DFT solver accommodating almost arbitrary boundary conditions. Therefore, evaluations of clusters and parent systems can use very similar algorithmic machinery, facilitating MLP training using mixed boundary condition datasets and providing a (mostly) apples-to-apples comparison when force matching (Section 2.1.1). \\

We employ the widely popular PBE GGA functional \cite{ref_perdew_generalized_1996} with Becke-Johnson D3 dispersion correction \cite{ref_johnson_post_hf_2006} and set a target grid spacing $h$ of 0.175 \AA{}. For periodic systems, the computational grid needs to fit inside a fixed cell. GPAW sets a value closest to the provided $h$, often resulting in a $(h_1, h_2, h_3)$ triplet to account for every cell vector. For finite clusters, we can set an arbitrary orthorhombic bounding box around the molecular fragment and the target $h$ can be matched exactly. Because GPAW imposes Dirichlet boundary conditions on the box faces, we surround clusters with a vacuum layer on every side. We found 4 \AA{}  to be an optimal width, producing results very similar to an infinite - very large - vacuum layer while keeping the grid relatively small. Similar conditions are enforced for the Poisson potential. GPAW has an `ExtraVacuumPoissonSolver' functionality that extends the grid for the electrostatic potential only. It is set to provide an additional Poisson vacuum layer of 8 \AA{} to avoid finite-size effects. In all calculations, k-point sampling is restricted to the Gamma point. We use finite difference stencils of maximum range and tri-heptic density interpolation. By default, GPAW computes formation energies, i.e., the energy of a structural configuration minus the vacuum energy of its constituent atoms. \\

Below, we discuss the main sources of error in GPAW single-point evaluations, as this is integral to correctly framing error metrics for trained MLPs. The functional approximation causes a systematic deviation from `true' QM molecular energies and forces. It is present in all calculations and can be disregarded if we take PBE-D3 as absolute ground truth. Numerical algorithms also involve errors of a stochastic nature. The computational grid introduces a dependence on $h$ and wrecks energy invariance and force equivariance under transformations of E(3). Random errors are caused by:
\begin{itemize}
    \item Grid mismatching. The grid spacing $h$ cannot always be freely chosen. Calculations with different $h$-values are effectively using different basis sets. Decreasing $h$ will monotonically converge the molecular energy, but  forces behave more spuriously. A dataset of periodic systems necessarily contains grid mismatches.
    \item The eggbox effect. Energy predictions follow a periodic variation under a translation of the system with regard to the grid, resembling a sinusoid with period $h$. Forces also vary with this period, although more erratically.
    \item Discrepancies in orientation. Rotating a structure will alter energies and forces, but a preferred orientation does not make sense. The error can often be avoided when comparing different structures, as orientation can be precisely controlled.
\end{itemize}

This numerical noise originates from the relative positioning of grid points and atomic nuclei. Its magnitude is governed by the overall grid spacing $h$ and the molecular system under investigation. 
We investigated stochastic force disparities using a dataset of 100 UiO-66(Zr) brick clusters. The dataset was reevaluated for (i) different grid spacings close to 0.175 \AA{}, (ii) various translations along cell vectors, and (iii) several arbitrary rotations. To distinguish contributing factors, only one of (i)-(iii) is varied at once. We found that each factor individually can lead to force discrepancies with a RMSE of 20 meV/\AA{} and maximal absolute errors on the order of 100 meV/\AA{} compared to a reference dataset. Under the crude approximation of independent random variables, the variances of these error sources combine constructively. Therefore, even if an MLP could interpolate the ground truth exactly, random noisy labels will still give rise to residual inference errors on test datasets. This analysis is not meant to discredit the validity of GPAW predictions, rather to set realistic expectations and get a sense of the DFT `noise floor'.

\section{NequIP architecture and training setup} \label{si_nequip}

We chose NequIP v0.5.6 \cite{ref_batzner_nequip_2022} as fundamental MLP architecture for all trained models. This section will discuss the most important neural network and training hyperparameters. If some setting is not specified, it is left as default. All MLPs will be made publicly available. \\

As explained in Section 4, we employed two network configurations in this work, which we call `base' (\textbf{mlp}$_\text{pr}$, \textbf{mlp}$^c_{\text{mix}}$, etc.) and `extended' (\textbf{mlp}$^c_{\text{sup}}$) in Table SI.\ref{si_table_nequip}, summarising the major differences between both parameter setups. \\

\begin{table}[h]
\centering
\includegraphics[width=\linewidth]{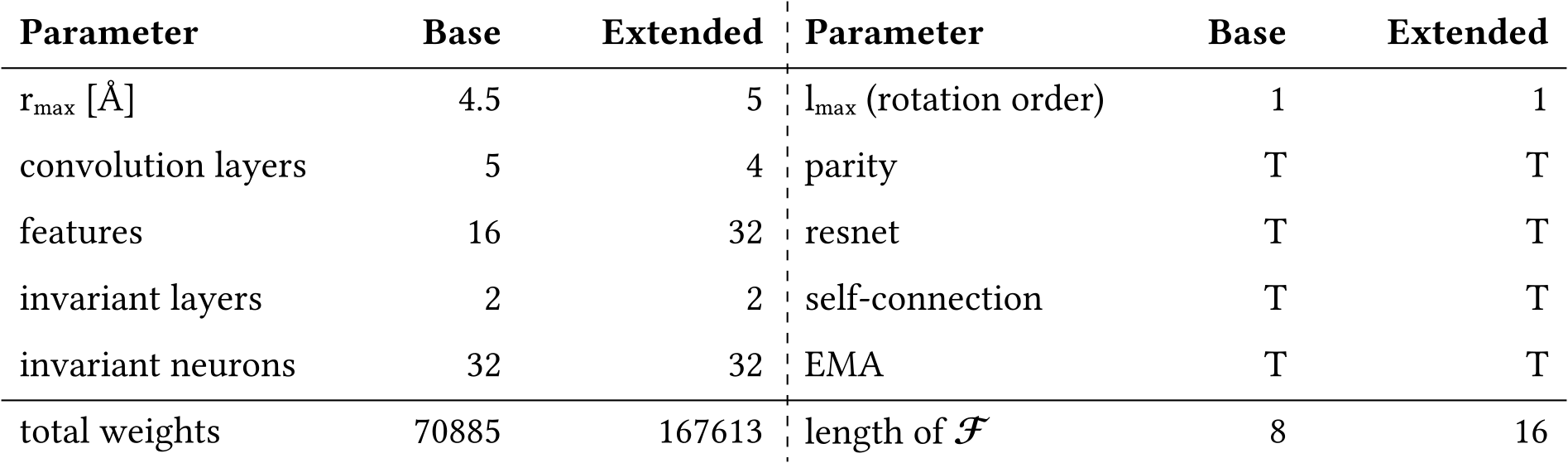}
\caption{Summary of NequIP hyperparameter setup.}
\label{si_table_nequip}
\end{table}

In terms of model training, we maintained an 80/20 training-validation split, randomly distributing configurations across both datasets. New models are initialised with trainable per-species scaling factors for energies and forces based on the training dataset force RMS. We utilise single precision, a mixed force and per-atom energy MSE loss function (respective weights 1 and 100), the default ADAM optimising scheme and the Pytorch `ReduceLROnPlateau' learning rate scheduler. MLPs are trained with a learning rate of 0.004 and a maximal batch size of 5, until a stopping condition is reached. Either:
\begin{itemize}
    \item the validation loss has decreased by less than 1e-5 over 250 epochs,
    \item the learning rate drops below 1e-5 by the scheduler, with a reducing factor of 0.75 and a patience of 25 epochs.
\end{itemize}

For `intermediate' models (during AL campaigns), we deviate slightly from these conditions, starting with a higher base learning rate and making the scheduler and early stopping thresholds more aggressive. This halts training prematurely and drastically shortens runtimes, without sacrificing much performance. Whenever the training dataset changes (once every AL cycle), we reset the stored exponential-moving-average (EMA) model and the optimiser momentum and give the model 10 warmup epochs before engaging the scheduler.

\section{Molecular dynamics} \label{si_md}

As mentioned in Section 3, we employ two MD software engines, depending on the type of ensemble that is sampled. Simulations in the isobaric-isothermal (NPT) ensemble are performed with OpenMM \cite{ref_eastman_openmm_2017}, whereas simulations in the fixed-volume NPT (N, V, $\bm{\sigma=0}$, T) ensemble use the in-house YAFF code. \cite{ref_verstraelen_yaff} For the latter, cell parameters are allowed to fluctuate in a way that preserves cell volume, and the dependent thermodynamic variable is the internal stress tensor. \cite{ref_rogge_comparison_2015} Table SI.\ref{si_table_md} summarises the main algorithmic components and parameters used in MD. Variables like temperature, pressure and simulation length are not mentioned; they vary across simulations (see main text). Every MD run initialises from a random seed and sets starting velocities according to a Maxwell-Boltzmann distribution.

\begin{table}[h]
\centering
\includegraphics[width=\linewidth]{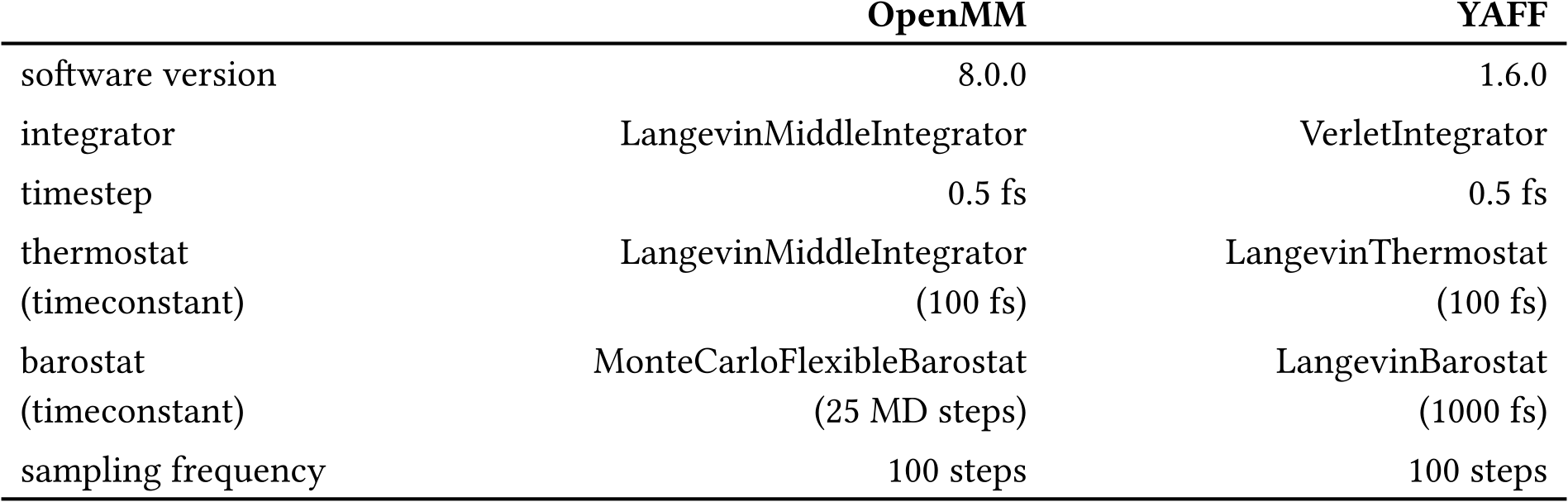}
\caption{A brief overview of the used MD simulation setup. For implementational details, we refer to the respective software documentation.}
\label{si_table_md}
\end{table}

\section{Active learning} \label{si_al}

This section reviews all components of the AL workflow and their interdependence as implemented in our cluster-based learning methodology (see Figure 2). \\

\textbf{Initalisation}\\
The user specifies a molecular system $S$ as a learning target - without size or boundary condition restrictions - and a set of thermodynamic state variables ENS in which the MLP will operate under inference. If available, a `seed' dataset $D^T_0$ containing some $\epsilon$ from $S(\epsilon)|_{\text{ENS}}$ can be provided to jumpstart the AL campaign, reducing overall runtime. Otherwise, we generate $D^T_0$ with random clusters extracted from spatially perturbed configurations of $S$. This data is used to train an initial MLP. \\

\textbf{Phase space exploration}\\
New structures are sampled through short, fixed-length MD walks at elevated temperatures to explore a diverse set of $\epsilon$, increasing model robustness and transferability. Walkers run in parallel. Early in the AL campaign, the MLP may wander into PES singularities, causing the run to explode. We mitigate these crashes through checkpointing and shortening simulation times in the first AL cycles. The final configuration of every walker is collected into a pool of sample structures. \\

\textbf{Data acquisition}\\
We find uncharted $\epsilon$ within the sample pool using the uncertainty estimation approach of Section 2.3. A feature density is parametrised based on the current iteration of $D^T(\epsilon)$, excluding $\epsilon$ in the validation set. The $N$ most valuable clusters are extracted from the pool of structures (see Section 2.4 and Section SI.\ref{si_extraction}). This process is computationally cheap and mostly negligible in the complete workflow. \\

\textbf{\textit{Ab initio} evaluation}\\
Labelling new data, i.e., freshly extracted clusters, is an embarrassingly parallel task. Generally, $D^T$ can contain finite and periodic structures. When dealing with mixed boundary conditions, special considerations are required to keep \textit{ab initio} computational settings (LOT, basis set, cutoff values) as consistent as possible (see Section SI.\ref{si_gpaw}). \\

\textbf{MLP (re)training}\\
New structures are randomly distributed over train and validation subsets of $D^T$, following a fixed splitting fraction. We retrain the existing MLP with all data. As the rate of model improvement slows down throughout epochs, we terminate training early for intermediate models by enforcing aggressive stopping conditions (see Section SI.\ref{si_nequip}). This avoids the asymptotic tail of the training curve, saving GPU time while sacrificing little accuracy. In the final AL cycle, we relax all early stopping conditions to extract maximal MLP performance. \\

The algorithmic extension towards a transferable model that learns multiple systems $S$ simultaneously is straightforward. We chose batched data sampling as opposed to an online sampling policy - i.e., monitoring MLP uncertainty during MD and terminating when some threshold is crossed - because it eliminates the need for an (arbitrary) threshold and allows direct data comparisons within a single batch to find the most interesting $\epsilon$. \\

Note the modular nature of this methodology. Every step can be adapted or replaced with numerous alternatives that perform a similar task, affording the user much freedom to tailor the implementation to a specific use case.

\section{Cluster extraction} \label{si_extraction}

Section 2.4 gave a brief outline of the idea behind cluster extraction. Here, we will thoroughly discuss the procedure, which consists of two steps: `core selection' and `fragment construction'. Assume we start with a trained MLP and a configuration of system $S$, label every $\epsilon$ with a density likelihood (see Figure 3) and want to create the most valuable cluster to incorporate in $D^T$, i.e., the cluster that results in the largest model improvement while remaining cost-effective. \\

\textbf{Core selection}\\
First, we decide which $\epsilon$ are most beneficial to extract, hence which atoms should form the core of the new cluster. Ideally, a core is spatially compact and contains many low-likelihood (high uncertainty) $\epsilon$. For MOFs, it makes sense to adhere to their natural building block composition (see Section SI.\ref{si_env_match}). After partitioning $S$ into a set of potential cores, we can rank each candidate using the likelihoods of its atoms. Currently, we select the core that includes the minimal likelihood of the entire structure, although more elaborate uncertainty-based heuristics can be devised. Within a core, one could e.g. sum all likelihoods below a specified threshold, compute the mean likelihood excluding any hydrogen atoms, or penalise its size, volume, or number of atoms. Finetuning an ultimate expression that maximises data efficiency is outside the scope of this publication. \\

\textbf{Fragment construction}\\
To extract $\left( \epsilon_i \right)_\text{parent}$ for every atom $i$ in the chosen core, we build a minimal mantle of parent atoms around it, ensuring Equation 2 holds through force matching (Section 2.1.1). Finally, we create a suitable termination layer of hydrogen atoms to saturate any dangling bonds at the cluster surface. Section SI.\ref{si_env_match} puts this recipe into practice, designing molecular fragments for the metal brick of UiO-66. Naturally, we do not want to repeat force matching for every structure containing interesting $\epsilon$. Once a suitable cluster blueprint - core, mantle and termination - has been identified, it can be reused across multiple configurations of $S$, provided no major structural changes occur (such as amorphisation, severe topology rearrangements, etc.). Generally, it is useful to establish generic cluster design rules for $S$ before starting the AL campaign. \\

Note that outer atoms in the mantle of molecular fragments will never replicate any $\left( \epsilon \right)_\text{parent}$ due to its finite radius. One could argue that $\epsilon$ which are not environment matched - defying Equation 2 - should not be included in $D^T(\epsilon)$ (i.e., through some masking feature). Notwithstanding, they still contain viable quantum mechanical information that could improve MLP inference. During training, extracted clusters are treated as regular non-periodic systems, and no distinction is made between atoms in the core or mantle.

\section{Example force matching} \label{si_env_match}

This section investigates how one can design and extract suitable clusters from a parent structure by following the force matching approach (Section 2.1.1). To ensure both $\bm{F}_i^\text{dft}$ and $\bm{F}_i^\text{mlp}$ fulfill Equation 3 for every core atom $i$, we must determine the interaction ranges of $\epsilon_i^\text{dft}$ and $\epsilon_i^\text{mlp}$. \\

The spatial extent of $\epsilon^\text{dft}$ is inherently determined by (the limitations of) the PBE D3 functional approximation and is a priori unknown. Conversely, $\epsilon^\text{mlp}$ is limited by the interaction radius $r_\text{max}$ of the model, determining how far the MLP can `look ahead'. If this hyperparameter - hard-coded in feature basis functions for Behler-like networks or message-passing layers in convolutional neural networks such as NequIP \cite{ref_behler_perspective_2016} -
is too small, the model cannot properly learn reference $\epsilon^\text{dft}$, limiting attainable accuracy. However, $r_\text{max}$ can be chosen arbitrarily large; it only sets an upper bound on the actual interaction range of $\epsilon^\text{mlp}$. \\

\begin{figure}
\centering
\includegraphics[width=0.4\linewidth]{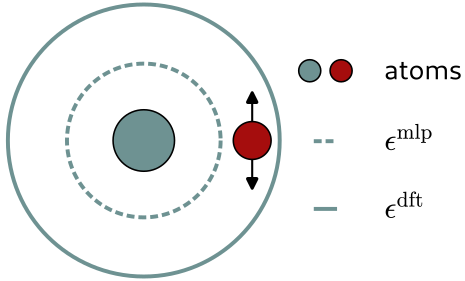}
\caption{Illustrating an environment mismatch between $\epsilon^\text{dft}$ and $\epsilon^\text{mlp}$.}
\label{si_fig_env_mismatch}
\end{figure}

Consider the situation depicted in Figure SI.\ref{si_fig_env_mismatch}, where $\epsilon^\text{dft}$ is more extensive than $\epsilon^\text{mlp}$. Moving the red atom alters $\epsilon^\text{dft}$ but not $\epsilon^\text{mlp}$, and the resulting variations in $\bm{F}^\text{dft}$ will not be reflected by $\bm{F}^\text{mlp}$. Therefore, if the MLP can perfectly reproduce the \textit{ab initio} ground truth, we expect that $\epsilon^\text{dft} = \epsilon^\text{mlp}$. Practically, discrepancies will inevitably persist and $\epsilon^\text{dft} \approx \epsilon^\text{mlp}$ for well trained models. \\

We try to define appropriate clusters for a UiO-66(Zr) unit cell with a linker defect ($S_\text{ld}$ in the main text) according to Section SI.\ref{si_extraction}. \\

\textbf{Core selection}\\
We divide $S_\text{ld}$ into loosely connected building blocks to form potential cluster cores. By breaking the C-C sigma bond at each carboxylic acid group, the unit cell deconstructs into two [Zr\textsubscript{6}O\textsubscript{4}(OH)\textsubscript{4}(CO\textsubscript{2})\textsubscript{12}] `bricks', two [Zr\textsubscript{6}O\textsubscript{4}(OH)\textsubscript{4}(CO\textsubscript{2})\textsubscript{11}(CO\textsubscript{2}H)] `defective bricks' - with coordination number 11 - and 23 [C\textsubscript{6}H\textsubscript{4}] `linkers'. Note that our decomposition differs from the conventional definition of zirconium bricks and BDC ligands. Suppose we want to extract some unknown $\epsilon$ in a defective brick. What atomic cluster is suited for this task? \\

\begin{figure}
\centering
\includegraphics[width=\linewidth]{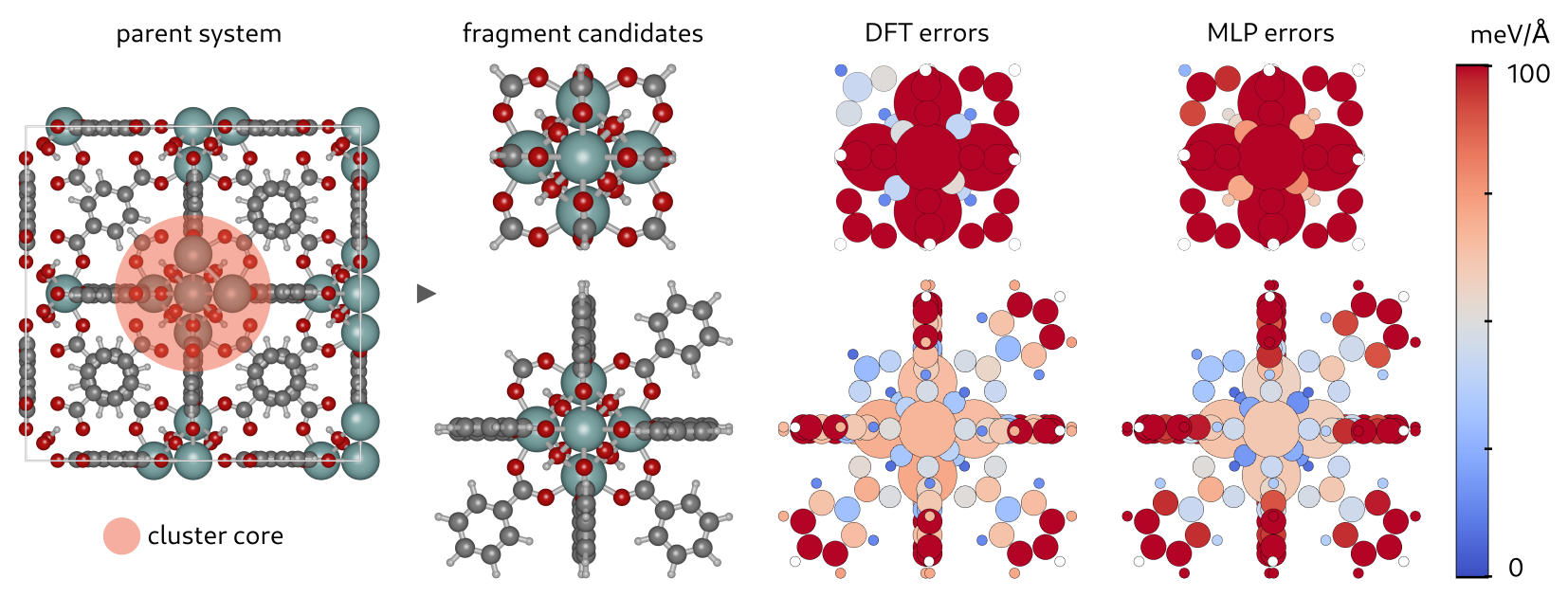}
\caption{Force matching for a defective brick in $S_\text{ld}$. The colour scale represents the per-atom RMSE between $\left( \bm{F} \right)_\text{parent}$ and $\left( \bm{F} \right)_\text{cluster}$ computed over $D_\text{ld}$. Terminating hydrogens are not coloured; they have no periodic counterpart.}
\label{si_fig_force_match}
\end{figure}

\textbf{Fragment construction}\\
The smallest possible candidate is simply the brick terminated with 11 hydrogens. A bigger fragment also includes the 11 neighbouring linker blocks as cluster mantle. To go larger, we must incorporate periodic duplicates of parent atoms, creating a fragment with more atoms than the original $S_\text{ld}$. That defeats the point entirely, so we only consider two cluster blueprints in the force matching procedure (see Figure SI.\ref{si_fig_force_match}), named `small' and `large' respectively. \\

\begin{table}[b]
\centering
\includegraphics[width=.5\linewidth]{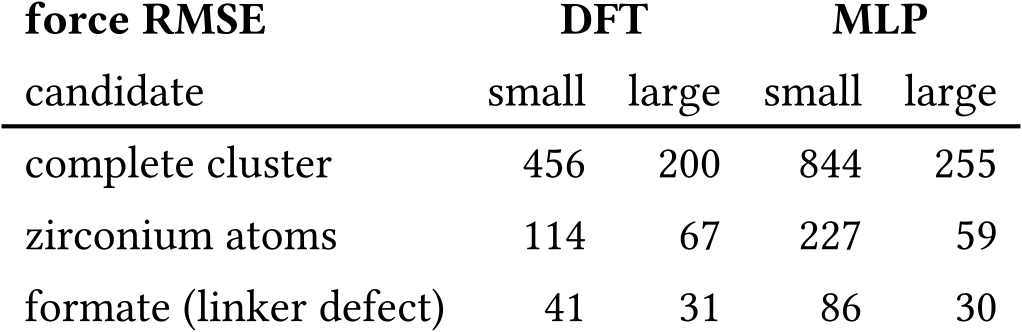}
\caption{Force deviation metrics in meV/\AA{} to complement Figure SI.\ref{si_fig_force_match}.}
\label{si_table_force_match}
\end{table}

\textbf{Force matching}\\
For a sample configuration of $S_\text{ld}$, we extract a cluster blueprint and contrast $\left( \bm{F} \right)_\text{parent}$ with $\left( \bm{F} \right)_\text{cluster}$ using the same LOT. We perform this comparison for every periodic structure in $D_\text{ld}$ and compute a per-atom RMSE of parent-cluster force discrepancies to achieve robust statistics. Figure SI.\ref{si_fig_force_match} shows these results for both `small' and `large' clusters, as well as DFT and MLP (\textbf{mlp}$_\text{pr}$) LOTs. Table SI.\ref{si_table_force_match} aggregates force errors over specific sets of atoms. \\

As a general trend, DFT seems more forgiving than \textbf{mlp}$_\text{pr}$ regarding $\epsilon$-mismatches, with the latter showing much larger RMSE values. Spurious extrapolation for out-of-dataset $\epsilon$ in clusters could explain the observed sensitivity of \textbf{mlp}$_\text{pr}$. DFT (MLP) deviations for the `small' blueprint average 114 (227) meV/\AA{} for Zr atoms and are even worse for outer [H-CO\textsubscript{2}] carboxylate anions. The only exception is the formate group corresponding to the linker defect. Because `small' clusters do not have any mantle to pad core atoms, they fail at capturing most $\epsilon$ from $S_\text{ld}$. We reach a different conclusion for the `large' blueprint. Here, force discrepancies are much lower for core atoms, averaging 67 (59) meV/\AA{} for Zr atoms and 31 (30) for the linker defect formate group with DFT (MLP) LOT. Only atoms near the cluster surface show hefty RMSE numbers, indicating that `large' clusters are superiour fragment candidates. \\

We did not identify any cluster blueprint that leads to exact force matching. Nevertheless, considering our discussion of the DFT noise floor (Section SI.\ref{si_gpaw}), we found a cluster blueprint for the defective brick in $S_\text{ld}$ in which $\epsilon^\text{dft} \approx \epsilon^\text{mlp}$ mostly holds. Repeating this exercise for the regular brick or linker block leads to similar results. As a general design rule, we posit that suitable clusters in (disordered) UiO-66-derived frameworks are given by a central core block surrounded by its first neighbours. This assumption is validated in the main text (see Table 3 and Table 4) and Table SI.\ref{si_table_metrics_full}: cluster-based MLPs can indeed describe periodic systems including various types of spatial disorder.

\section{MLP uncertainty and force errors} \label{si_uncertainty}

\begin{figure}
    \centering
    \includegraphics[width=0.9\linewidth]{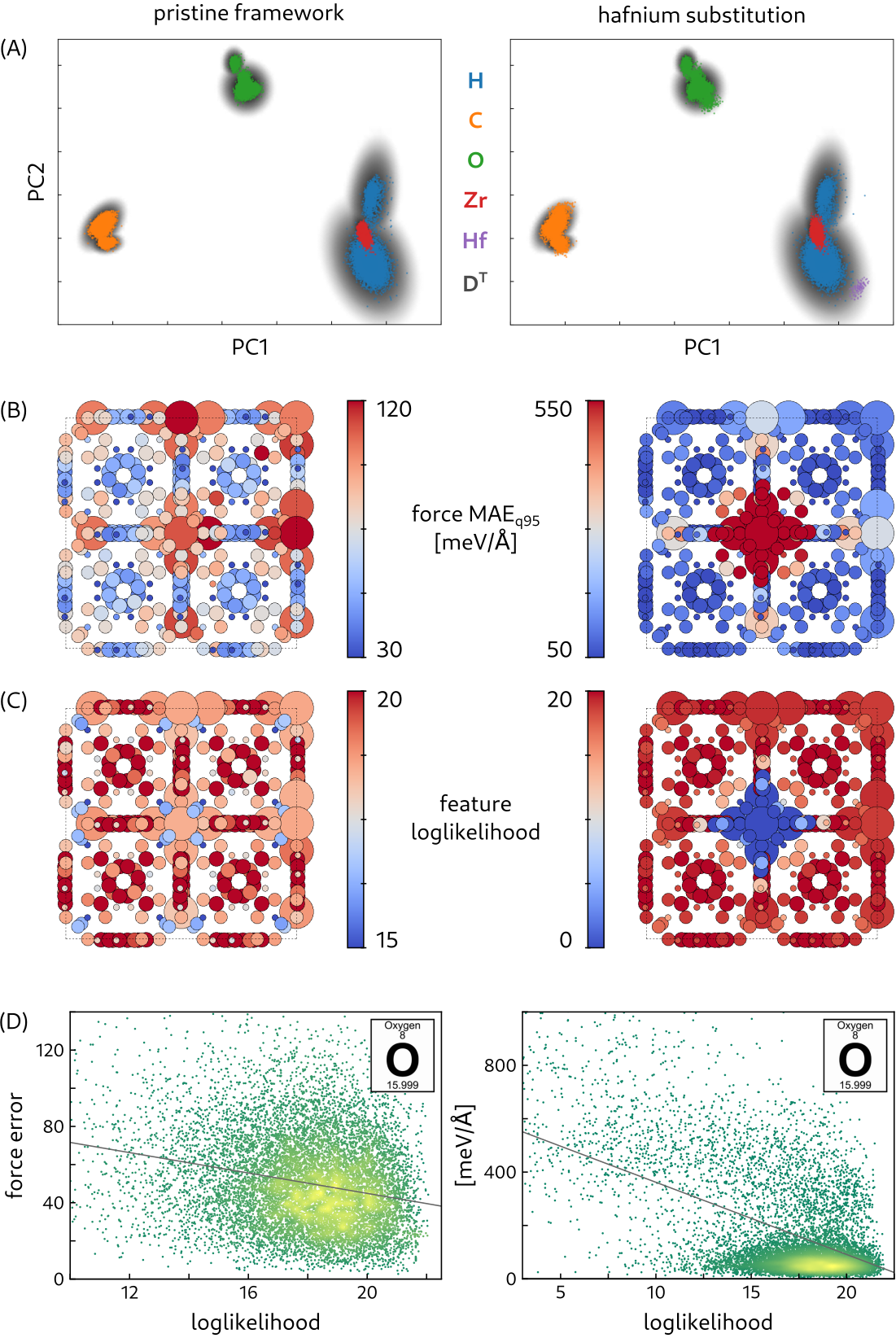}
    \caption{Comparing \textbf{mlp}$_\text{pr}$ inference on $D_\text{pr}$ (left) and $D_\text{hf}$ (right). (A) Feature space representation of all $\epsilon$, where $\bm{\mathcal{F}}_i$ is colour-coded according to the element of atom $i$. The grey shading shows a GMM density fit to $D^T_\text{pr}$. (B) Per-atom force $\text{MAE}_{P95}$ of \textbf{mlp}$_\text{pr}$. (C) Per-atom average loglikelihood of $\epsilon$ based on a feature density fit of $D^T_\text{pr}$. (D) Loglikelihood of $\epsilon_i$ versus $\left| \bm{F}_i^\text{mlp} - \bm{F}_i^\text{dft} \right|$ for oxygen atoms in $D_\text{pr}$ and $D_\text{hf}$. Colour is indicative of scatter density.}
    \label{si_fig_uncertainty}
\end{figure}

In Section 2.3 we represent $D^T(\epsilon)$ using a density distribution fitted to $\bm{\mathcal{F}}$-descriptors in MLP feature space. The underlying hypothesis is that - after model training - MLP uncertainties (and inference errors) inversely correlate with density likelihoods. We test this premise in Figure SI.\ref{si_fig_uncertainty} by analysing force error metrics of \textbf{mlp}$_\text{pr}$ evaluated on $D_\text{pr}$ (left) and $D_\text{hf}$ (right) in relation to the feature contents of each respective dataset. \\

Figure SI.\ref{si_fig_uncertainty}.A shows a feature space representation of $D_\text{pr}(\epsilon)$ and $D_\text{hf}(\epsilon)$, where each point $\bm{\mathcal{F}}_i$ is colour-coded according to the element of atom $i$. In grey, we have superimposed a GMM density fit to $D^T_\text{pr}$. Keep in mind that $\bm{\mathcal{F}}$-vectors for \textbf{mlp}$_\text{pr}$ are originally 8-dimensional. Figure SI.\ref{si_fig_uncertainty}.A is a 2D projection on the first principal components of $D^T_\text{pr}$ in $\bm{\mathcal{F}}$-space. It illustrates how \textbf{mlp}$_\text{pr}$ separates features by atomic element. Additionally, the training density of $D^T_\text{pr}$ (grey) overlaps nicely with $D_\text{pr}$ test points, indicating the datasets contain similar $\epsilon$. This is an expected result; they both consist of configurations of $S_\text{pr}$. A significant mismatch between $D^T_\text{pr}$ and $D_\text{pr}$ would point towards incomplete or incorrect sampling in either dataset. The overlap is much worse for $D_\text{hf}$: a new hafnium $\bm{\mathcal{F}}$-cloud appears and the spread on C and O features is more pronounced. Using Figure SI.\ref{si_fig_uncertainty}.A, we can visually confirm that $D_\text{hf}(\epsilon)$ contains $\epsilon \notin D^T_\text{pr}(\epsilon)$. \\

Figure SI.\ref{si_fig_uncertainty}.B and Figure SI.\ref{si_fig_uncertainty}.C depict the per-atom force $\text{MAE}_{P95}$ and average $\bm{\mathcal{F}}$-loglikelihood for $D_\text{pr}$ and $D_\text{hf}$ using \textbf{mlp}$_\text{pr}$. Here, the likelihood of atom $i$ is computed from a density fit to the $\bm{\mathcal{F}}$-cloud of atoms matching in atomic number (e.g., only H atoms). Constructing one smaller GMM per atomic element is considerably cheaper than parametrising a single large GMM on all data, and it decouples each distribution, enabling more freedom to examine the $\epsilon$ of every element independently. Note that the range of predicted (log)likelihoods depends on $ D^T_\text{pr}(\epsilon)$ and the $\bm{\mathcal{F}}$-dimension of \textbf{mlp}$_\text{pr}$. In Figure SI.\ref{si_fig_uncertainty}.B, force errors are generally small for $D_\text{pr}$. The MLP is more accurate in linkers, which is not surprising given the fraction of $\epsilon$ in $D^T_\text{pr}$ that corresponds with bricks ($\pm16\%$). Figure SI.\ref{si_fig_uncertainty}.C provides a fairly symmetric and mundane likelihood distribution. The average likelihood of e.g., hydrogen in bricks is vastly lower than hydrogen in linkers, coinciding with the relative occurrence of both types of H-atoms (16 vs 96 per unit cell). This again follows expectation and underlines the similarity between $D_\text{pr}$ and $D^T_\text{pr}$. However, the analysis differs strongly for $D_\text{hf}$. Figure SI.\ref{si_fig_uncertainty}.B shows enormous inference errors near the hafnium substitution, mirrored by a large drop in average likelihoods in Figure SI.\ref{si_fig_uncertainty}.C. These outliers drown out any small deviations in the remaining cell. On a visual basis, we can predict large force errors in $D_\text{hf}$ by the corresponding density likelihood. \\

Finally, Figure SI.\ref{si_fig_uncertainty}.D plots the force error $\left| \bm{F}_i^\text{mlp} - \bm{F}_i^\text{dft} \right|$ versus the $\bm{\mathcal{F}}$-loglikelihood of atom $i$, only including oxygen atoms for readability. If our hypothesis holds, we should find a negative correlation between both quantities. At first glance, the scatter plot for $D_\text{pr}$ seems mostly noise without emerging trends. Nevertheless, we find a Pearson correlation coefficient of $-0.26$ and perform a least-squares linear fit that returns a negative slope and $R^2$ value of $0.07$. On average, lower likelihoods indeed correspond with larger force errors, but the linear fit only explains a small fraction of the total data variance. As a result, the likelihood is a rather poor predictor of force error for a single sample $\epsilon$ in $D_\text{pr}$ - since all atomic interactions are already contained in $D^T_\text{pr}$ to some extent and errors are relatively low. On the contrary, the scatter plot for $D_\text{hf}$ has noticeably more structure. This comparison results in a Pearson coefficient of $-0.84$ and a linear trend with $R^2$ of $0.70$. Our likelihood-based approach is much more predictive for $D_\text{hf}$, as it contains $\epsilon \notin D^T_\text{pr}(\epsilon)$. Note that we have no basis to assume linear behaviour between $\bm{\mathcal{F}}$-loglikelihoods and force errors; it purely establishes a trendline. Analogous figures for the remaining elements (H, C, Zr) lead to qualitatively similar outcomes. \\

Overall, we conclude that the assumed correlation between MLP inference errors and feature density 
likelihoods certainly exists and becomes more outspoken for $\epsilon$ not in the model's training dataset, which are precisely the $\epsilon$ we want to identify and extract.

\section{MLP cross-validation} \label{si_crossval}

\begin{table}[]
\centering
\includegraphics[width=.8\linewidth]{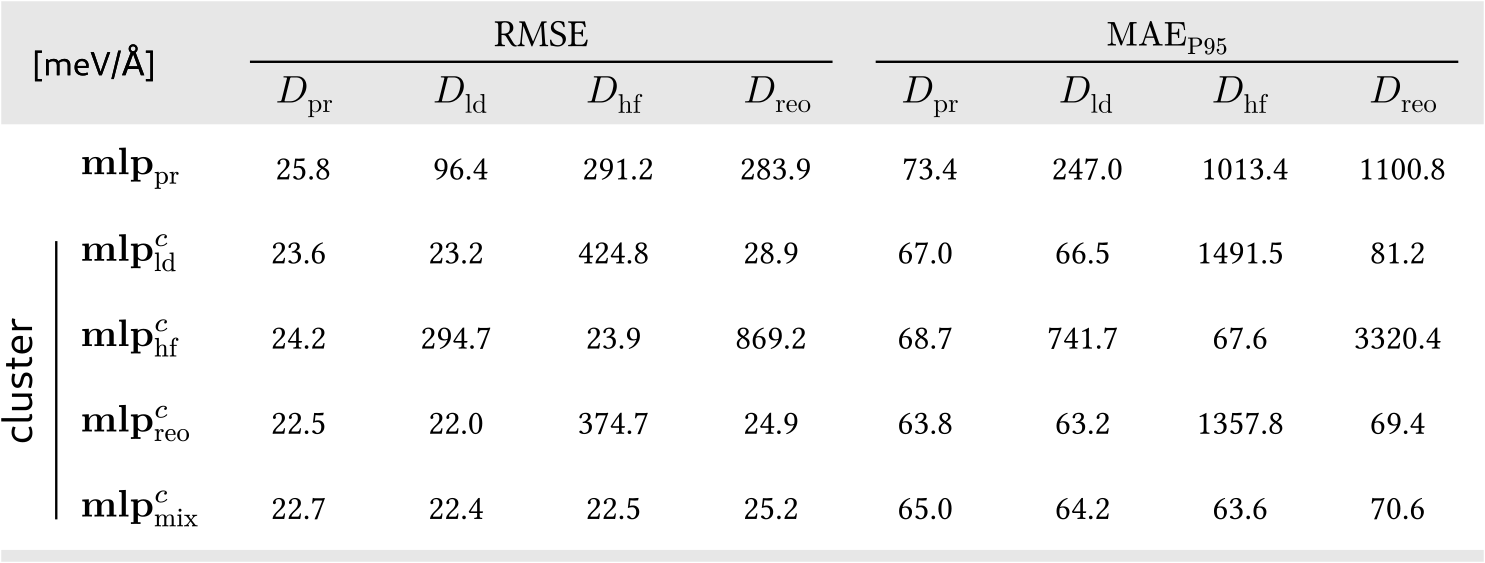}
\caption{Cross-validation of force RMSE and $\text{MAE}_{P95}$ metrics for various MLPs (rows) and test datasets (columns). All values are given in meV/\AA{}.}
\label{si_table_val_f}
\end{table}

In Section 4.2, we have discussed each point defect (ld, hf and reo) independently.
In the following paragraphs, we will cross-validate MLP metrics across test datasets to uncover hidden relations between different kinds of spatial disorder.
The analysis includes four periodic models - \textbf{mlp}\textsubscript{pr}, \textbf{mlp}\textsubscript{ld}, \textbf{mlp}\textsubscript{hf} and \textbf{mlp}\textsubscript{reo} - and four cluster models.
We select the final MLP from every learning curve in Figure 5.B ($N = 500$) and train one additional model by merging all training data, i.e., $D^T_{\text{pr}}$ and 500 fragments of each type (see Table SI.\ref{si_table_models}).
To remove ambiguity, these are named \textbf{mlp}$^c_{\text{ld}}$, \textbf{mlp}$^c_{\text{hf}}$, \textbf{mlp}$^c_{\text{reo}}$ and \textbf{mlp}$^c_{\text{mix}}$.
A superscript $c$ indicates the MLP is trained using clusters, alongside the basic $D^T_{\text{pr}}$ dataset.
Table SI.\ref{si_table_val_f} and Table SI.\ref{si_table_val_e} report atomic force and molecular energy metrics, using RMSE and $\text{MAE}_{P95}$ or $\Delta E_{\text{avg}}$ and $\Delta E_{\text{std}}$ respectively (see Section 3).
For force errors, we limit periodic models to \textbf{mlp}\textsubscript{pr}, which suffices to examine emerging trends.
Table SI.\ref{si_table_metrics_full} provides a full overview of all models and test sets. \\

\underline{Forces:} Table SI.\ref{si_table_val_f} features a strong dichotomy in error magnitudes.
At the low end, RMSE values bottom out around 20-30 meV/\AA{}, which corresponds to an $\text{MAE}_{P95}$ between 60-80 meV/\AA{}.
These errors near the convergence threshold of our DFT computations, meaning the MLP cannot extract more information from the numerical noise in its reference data.
Here, the predominant source of MLP inaccuracy, regarding the true QM ground truth, is the functional approximation. \cite{ref_behler_machine_2021}
At the high end, metrics often skyrocket by an order of magnitude, caused by a small number of deeply erroneous force predictions.
The inability to describe local interactions indicates shortcomings in the training data.
This is especially obvious for $D_{\text{hf}}$, where good model performance is only achieved when a training set includes Hf atoms.
Analogous conclusions follow for the other test cases. \\

Moreover, we uncover a remarkable reciprocal relationship between linker and node defects; only one is needed to describe either correctly.
While $S_{\text{ld}}$ and $S_{\text{reo}}$ both introduce formate capping groups to replace missing ligands, a linker defect creates two 11-coordinated bricks, whereas all bricks are 8-coordinated for $S_{\text{reo}}$.
These differences are not decisively reflected in Table SI.\ref{si_table_val_f} and hint at a large $\epsilon$-overlap within the two systems, reinforcing the idea that chemical environments have limited interaction radii.
Nevertheless, \textbf{mlp}$^c_{\text{reo}}$ slightly outperforms \textbf{mlp}$^c_{\text{ld}}$ on both test sets.
We speculate that $S_{\text{reo}}$-clusters carry more new information per structure than $S_{\text{ld}}$-clusters (1 vs 4 formate groups). \\

All cluster models marginally surpass \textbf{mlp}\textsubscript{pr} for pristine UiO-66; they are trained on a superset of $D^T_{\text{pr}}$.
However, \textbf{mlp}$^c_{\text{mix}}$ is not invariably the most accurate model, even though it has the largest dataset.
Learning three defect types simultaneously sacrifices some accuracy for any single defect. \\

\begin{table}[]
\centering
\includegraphics[width=.8\linewidth]{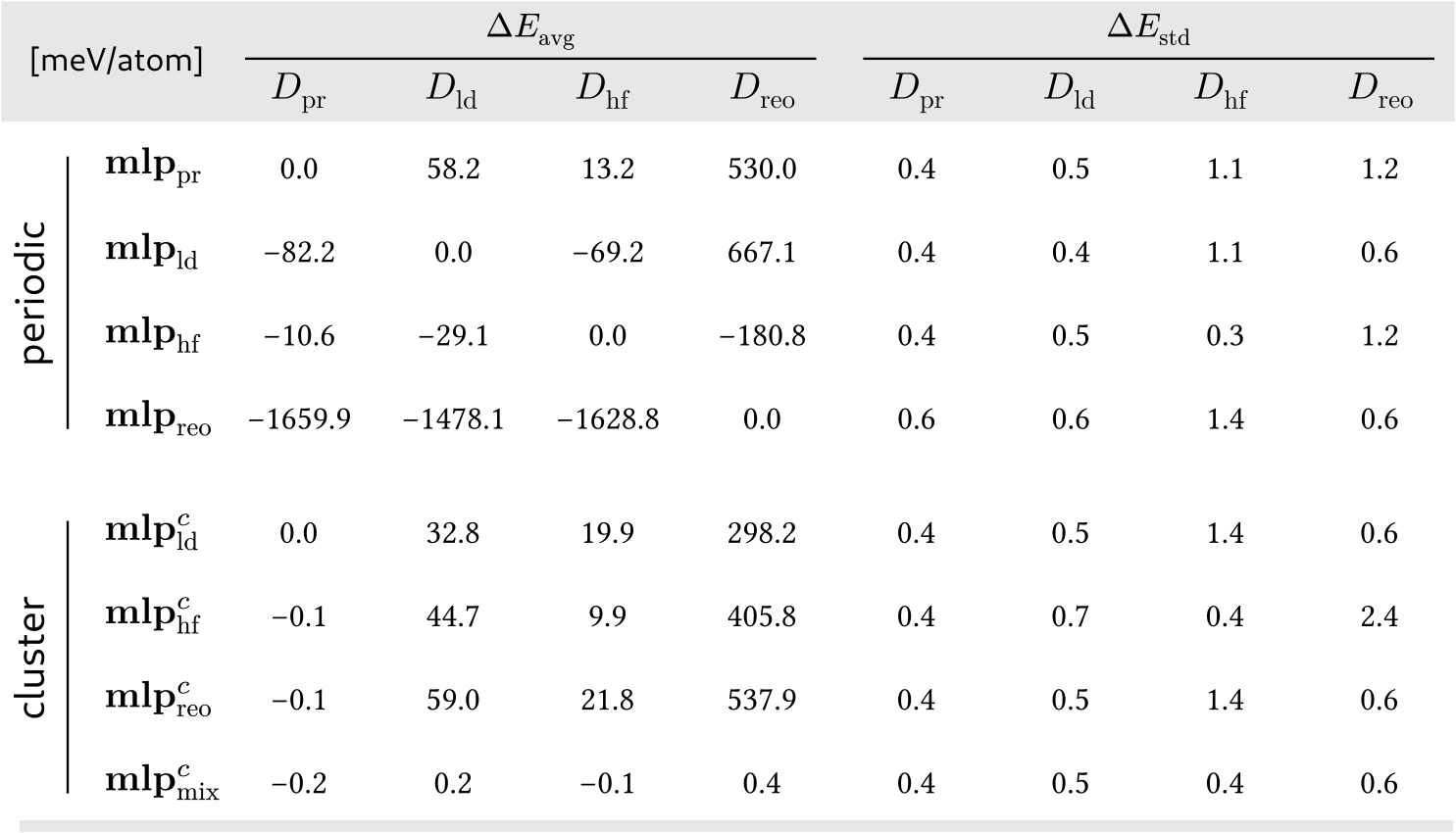}
\caption{Cross-validation of energy $\Delta E_{\text{avg}}$ and $\Delta E_{\text{std}}$ metrics for various MLPs (rows) and test datasets (columns). All values are given in meV/atom.}
\label{si_table_val_e}
\end{table}

\underline{Energies:} $\Delta E_{\text{avg}}$ and $\Delta E_{\text{std}}$ exhibit sizeable differences in scale (see Table SI.\ref{si_table_val_e}). While the latter is always of order 1 meV/atom, the former jumps around in a seemingly erratic manner. $\Delta E_{\text{std}}$ is consistently small for $D_{\text{pr}}$ and $D_{\text{ld}}$, despite large force errors near the missing linker for three out of eight models.
These do not propagate substantially into the total energy error, illustrating the contrast between local and global molecular properties.
MLP accuracy varies more strongly for $D_{\text{hf}}$, indicative of the large AOE of Hf substitutions.
This holds for $D_{\text{reo}}$ too.
As expected from Table SI.\ref{si_table_val_f}, the connection between ld and reo defects is also reflected in energy metrics.
Generally, accurate force predictions imply low $\Delta E_{\text{std}}$ values, but the inverse is not guaranteed. \\

For periodic models, $\Delta E_{\text{avg}}$ vanishes when training and test systems match, while every other combination results in moderate to huge energy offsets.
This phenomenon is inherent to MLP architectures.
The total energy $E$ is a sum of atomic energies $e$, and learning appropriate $e$ - apart from being non-physical - is a severely underdetermined problem.
Consider \textbf{mlp}\textsubscript{pr} and $S_{\text{pr}}$, which has eight C atoms for every Zr atom.
Whilst training, the model is completely free to reduce carbon $e$ by 10 eV and compensate by increasing $e$ for Zr atoms by 80 eV, as such shifts have no impact on $E$ or its derivatives.
In this example, $\Delta E_{\text{avg}}$ has not altered for $D^T_{\text{pr}}$ or $D_{\text{pr}}$, but it will change for every dataset with a different C/Zr ratio.
Excess degrees of freedom in parametrising $E$ explain the ostensibly random behaviour of $\Delta E_{\text{avg}}$.
They can be constrained by including structures with diverse elemental compositions during MLP training, as thoroughly shown in \cite{ref_eckhoff_molecular_2019}.
In full agreement, Table SI.\ref{si_table_val_e} shows that combining the three types of clusters with $D^T_{\text{pr}}$ results in low $\Delta E_{\text{avg}}$ values across the board.
Datasets limited to just one cluster type leave too much freedom and do not lead to a systematic improvement over \textbf{mlp}\textsubscript{pr}.
Only \textbf{mlp}$^c_{\text{mix}}$ reliably predicts accurate molecular energies for every UiO-66 variant. \\

From this discussion, we conclude that missing $\epsilon$ can be detected through force metrics, but not consistently through energy errors.
Different types of disorder can introduce comparable new $\epsilon$ (i.e, $S_{\text{ld}}$ and $S_{\text{reo}}$).
Finally, the most transferable model is trained from the most diverse training dataset.

\section{Mechanical characterisation} \label{si_props}

We characterise the mechanical properties of different frameworks through pressure-versus-volume (PV) and energy-versus-volume (EV) curves. Below, we provide technical details regarding the simulations involved and some results referred to in the main text.

\subsection{Computing PV curves} \label{si_props_pv}

PV profiles are derived from MD simulations at finite temperatures. We use NPT simulations in the elastic strain regime to find the equilibrium volume under applied pressure $\langle V(P_\text{ext}) \rangle$. Elsewhere, stochastic barostat fluctuations may cause premature phase transitions, and we switch to the (N, V, $\bm{\sigma=0}$, T) ensemble, which constrains cell volume but allows its shape to vary freely \cite{ref_rogge_comparison_2015}, to find the average internal pressure at a fixed volume  $\langle P_\text{int} (V) \rangle$. Under equilibrium conditions, these ensembles should agree and data points can be combined to describe the full PV behaviour. This combined approach exploits the computational efficiency of OpenMM when NPT volume fluctuations are limited and converge easily, and the improved stability of YAFF near maxima or unstable branches of the PV curve. The bulk modulus $K$ at equilibrium volume $V_0$ is defined as 
\begin{equation}
K = - V \left. \frac{\partial P}{\partial V} \right|_{V_0}
\end{equation}

\begin{figure}
\centering
\includegraphics[width=0.75\linewidth]{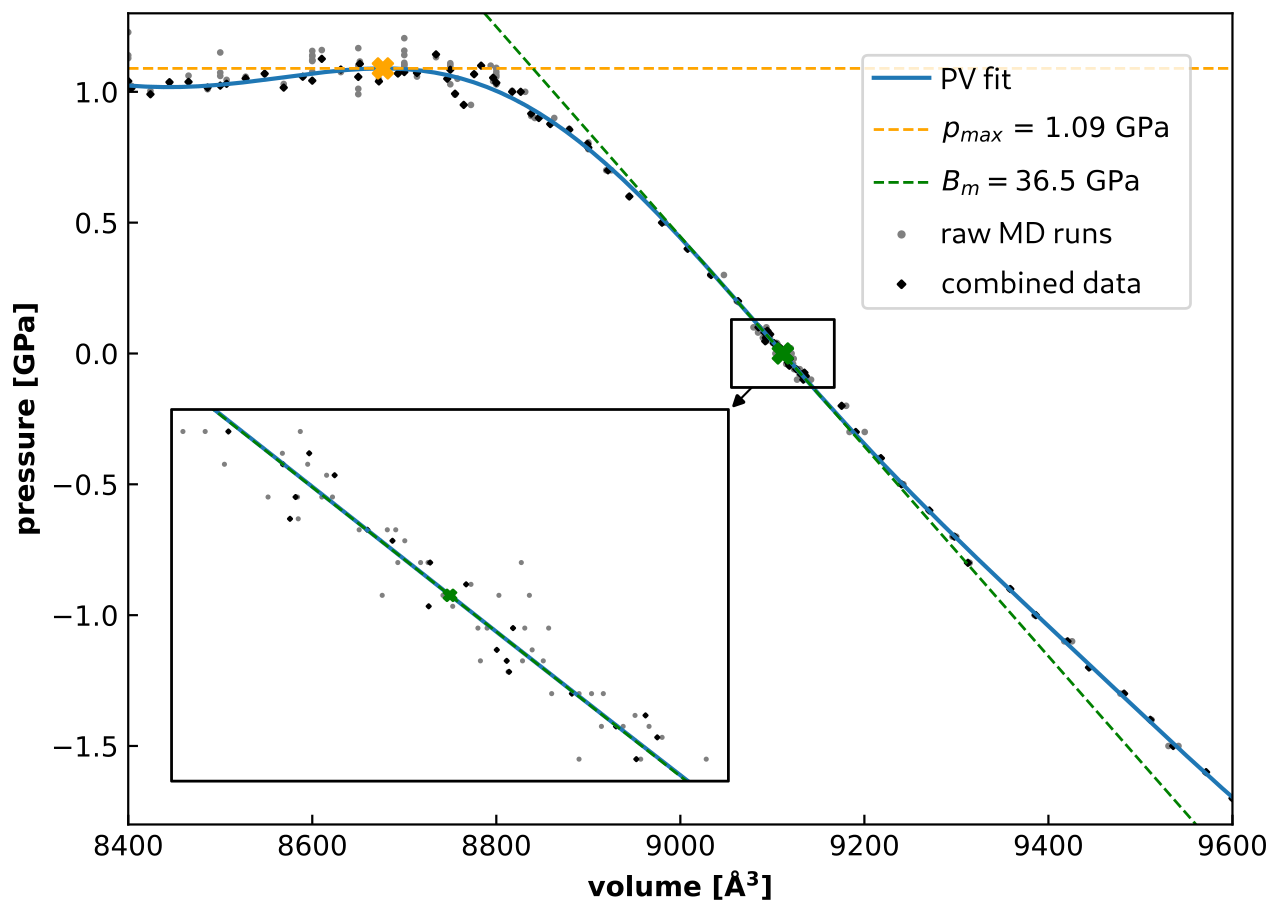}
\caption{Fitted PV curve for pristine UiO-66(Zr), derived using \textbf{mlp}$^c_{\text{sup}}$. Grey dots represent individual MD runs. Runs with identical controlled state variables are combined to provide better averages (black). The inset shows a zoom around vacuum pressure ($-100 \rightarrow 100$ MPa).}
\label{si_fig_pv_example}
\end{figure}

Simulations are performed for a grid of thermodynamic conditions. Each MD run initialises from an equilibrated structure and lasts roughly 50 ps (or longer, HPC walltime permitting), logging the energy, volume and (applied/internal) pressure every 50 fs. The first 20\% of recorded data is discarded, to allow for further system equilibration. By way of example, Figure SI.\ref{si_fig_pv_example} shows the simulation results and PV curve of pristine UiO-66 ($S_\text{pr}$), derived using \textbf{mlp}$^c_{\text{sup}}$. In this instance, NPT runs were performed for applied pressures between -1.8 GPa and 1.2 GPa, with an interval of 100 MPa. The spacing is reduced to 10 MPa around vacuum pressure (see inset), because the fitted curve should accurately capture the bulk modulus $K$. Between 8400-8900 \AA$^3$, we perform (N, V, $\bm{\sigma=0}$, T) simulations with a volume spacing of $\pm$ 25 \AA$^3$. The pressure and volume ranges probed depend on the material of interest; NPT sampling far above $P_\text{max}$ is wasted effort. In Figure SI.\ref{si_fig_pv_example}, different runs at identical thermodynamic conditions - e.g., two (N, P=0 Pa, T= 300 K) trajectories - show some spread on the final PV data, either due to insufficient sampling or due to simulations being restricted to separate regions of configuration space. For a better ensemble average, crucial simulations (around vacuum pressure and the PV maxima) are executed multiple times and their recorded data is combined. \\

The final PV profile is constructed using a nonparametric Gaussian Process implemented in scikit-learn. \cite{ref_pedregosa_sklearn_2011} We assess the convergence of each curve by varying the percentage of simulation data used during fitting, i.e., using 70\% of all data means discarding the first 30\% of every MD trajectory. If material properties remain approximately constant over a range of 40\% to 80\% of data used, we consider the curve to be converged. Otherwise, additional simulations are performed. We found $P_\text{max}$ to be more robust than $K$, and decided on a threshold of $15$ MPa and 1 GPa, respectively. Note that the values of Table 5 could vary slightly depending on the analysed data fraction and subsequent rounding. These fluctuations are small compared to the discrepancies one might observe when computing PV curves using different MLPs, training datasets or LOTs.

\subsection{Computing EV curves} \label{si_props_ev}

EV profiles are computed at 0 K through structure optimisations using algorithmic solvers from the `Atomic Simulation Environment' (ASE v3.22.1). \cite{ref_ase_2017} The procedure is outlined in Ref. \cite{ref_vanpoucke_mechanical_2015}: (i) scale the system of interest over a grid of volume points near its vacuum volume, (ii) perform a fixed-volume optimisation at every volume point, allowing the cell shape and atomic positions to relax and (iii) fit an equation-of-state (EOS) to the resulting data points to describe the $E(V)$ relation. A bulk modulus $K$ can be computed from its second derivative. \\

\begin{figure}
\centering
\includegraphics[width=0.75\linewidth]{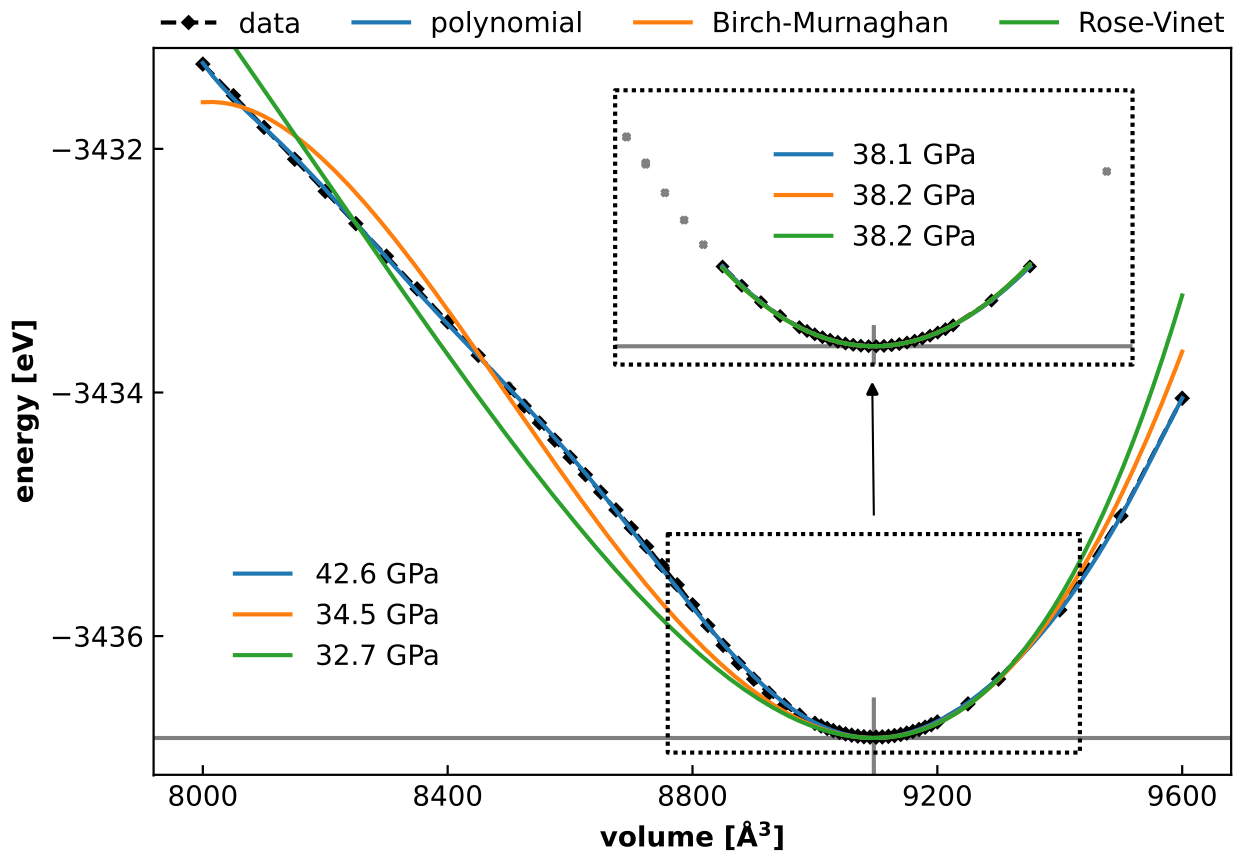}
\caption{EV profiles for pristine UiO-66, computed with \textbf{mlp}$^c_{\text{sup}}$. Depending on the EOS used and the volume interval fitted, different $K$ values are found.}
\label{si_fig_ev_uio66}
\end{figure}

We settled on a maximal force tolerance of 1 meV/\AA{} (per component) to decide when PES optima have been found. Literature proposes many EOSs; each with their particular strengths and weaknesses. One needs to decide which EOS to use and over what volume interval it should be fit. Figure SI.\ref{si_fig_ev_uio66} shows different EV profiles for UiO-66, demonstrating that both choices should be considered carefully. In this example, we parametrise a simple polynomial, a Birch-Murnaghan EOS \cite{ref_birch_finite_1947} and a Rose-Vinet EOS \cite{ref_vinet_compressibility_1987}. Over the volume range of 8000-9600 \AA$^3$, only the polynomial has the functional flexibility to fit every EV point accurately and each EOS leads to a different value of $K$. By reducing the interval of volume points considered in the fit (see inset), the ensemble of EOSs converges to a single curve with one unique bulk modulus, which we take to be representative. Every $K$ reported in Section 4.4 is derived using this criterion of EOS agreement.

\subsection{PV curves for double linker defects} \label{si_pv_results_ld2}

\begin{table}[]
\centering
\includegraphics[width=\linewidth]{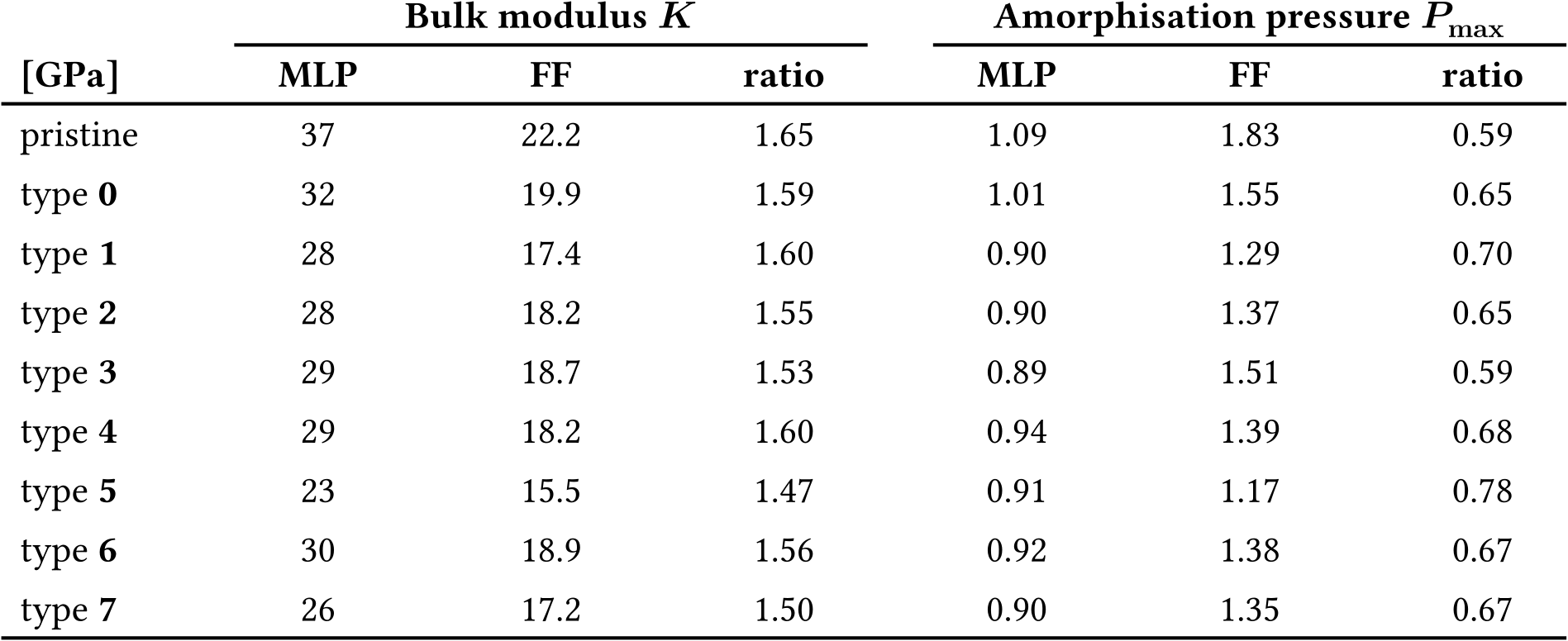}
\caption{Bulk moduli and amorphisation pressures for (defective) UiO-66 unit cells with up to two linker defects. MLP values correspond to the \textbf{mlp}$^c_{\text{sup}}$ model of the main text, FF values were derived through system-specific force fields \cite{ref_rogge_thermodynamic_2016}. All values (except ratios) are in GPa.}
\label{si_table_ld2}
\end{table}

\begin{figure}
\centering
\includegraphics[width=0.75\linewidth]{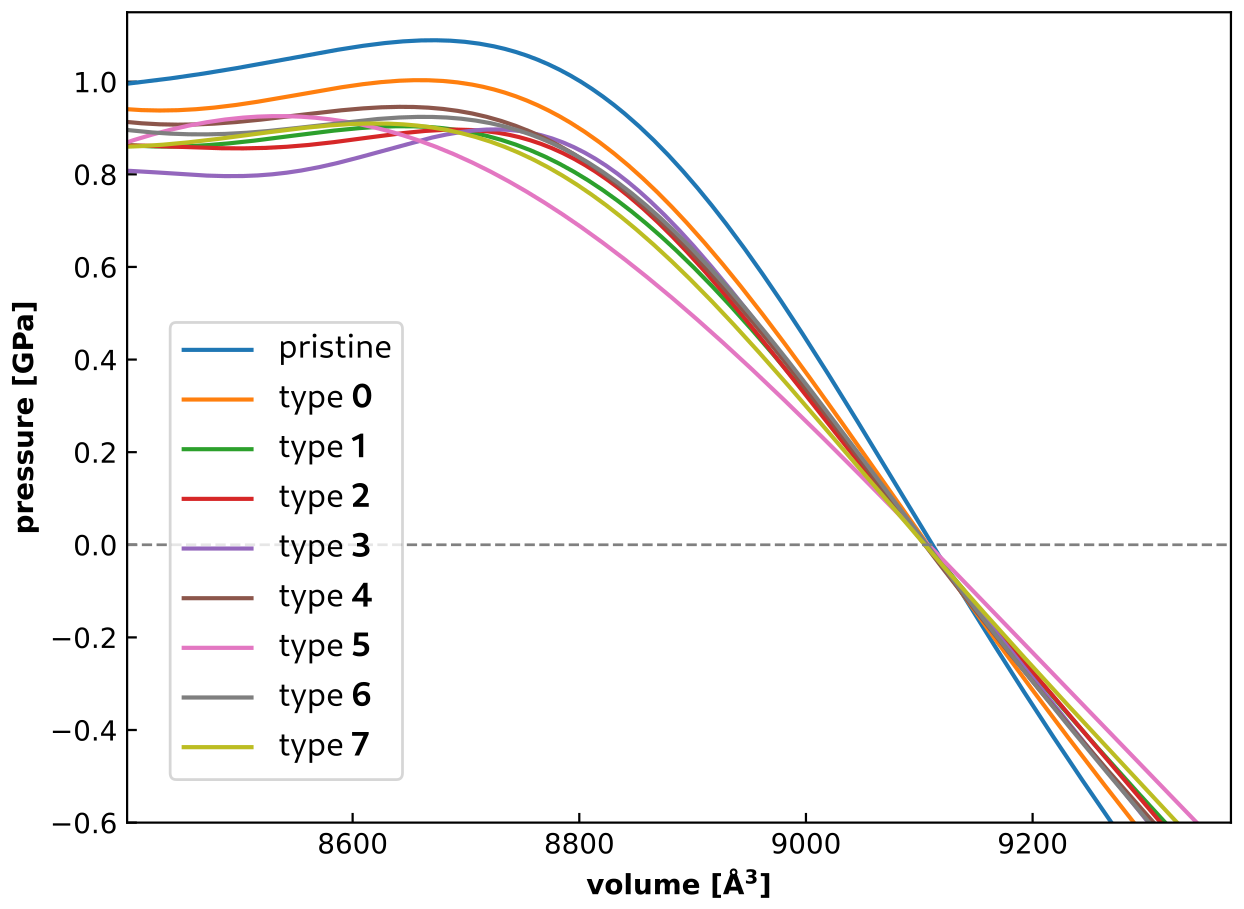}
\caption{Fitted PV curves for (defective) UiO-66 unit cells with up to two linker defects, derived using \textbf{mlp}$^c_{\text{sup}}$.}
\label{si_fig_pv_ld2}
\end{figure}

Figure 7 and Table 5 of the main text show PV curves and derived properties for every unit cell system considered in this work, but lump cells with double linker defects together. Here, we provide a more fine-grained analysis for the seven `ld-2' systems ($S_\text{ld}^{1-7}$, see Figure SI.\ref{si_fig_double_linker_defect}) and compare with earlier force field (FF) results by Rogge et al. \cite{ref_rogge_thermodynamic_2016} For reference, we also include the pristine UiO-66 cell $S_\text{pr}$ and the variant with a single linker defect $S_\text{ld}$, while adopting the nomenclature from Ref. \cite{ref_rogge_thermodynamic_2016} (see Table SI.\ref{si_table_ld2} and Figure SI.\ref{si_fig_pv_ld2}). \\

We find a systematic underestimation of FF bulk moduli compared to our \textbf{mlp}$^c_{\text{sup}}$ results, but the ratio is relatively consistent. A Pearson correlation coefficient of $0.99$ and a trendline with $R^2 = 0.98$ indicate a robust linear relation between both levels of theory.
This is likely because bulk moduli in MOFs are largely determined by overall geometry and topology under equilibrium conditions - which FFs can accurately describe - whereas precise atomic interactions - where FFs tend to struggle - only take a secondary role. \cite{ref_moghadam_structure_mechanical_2019}
Should this relation hold in general, we could stick to (cheaper) FF descriptions of systems and extrapolate the corresponding MLP bulk modulus without having to train new models. \\

The story is different for $P_\text{max}$. Here, MLP results are always significantly lower than FF predictions. A Pearson correlation coefficient of 0.82 and linear fit with $R^2 = 0.68$ show that both 
LOTs no longer show a clear trend. In particular, the relative order of `ld-2' systems is mostly lost. However, $P_\text{max}$ represents a critical point on the PV curve, and the FFs in Ref. \cite{ref_rogge_thermodynamic_2016} were parametrised at $V_0$. Unsurprisingly, model agreement is better for $K$ than $P_\text{max}$.

\subsection{EV curves for various topologies} \label{si_ev_results}

\begin{figure}
\centering
\includegraphics[width=0.9\linewidth]{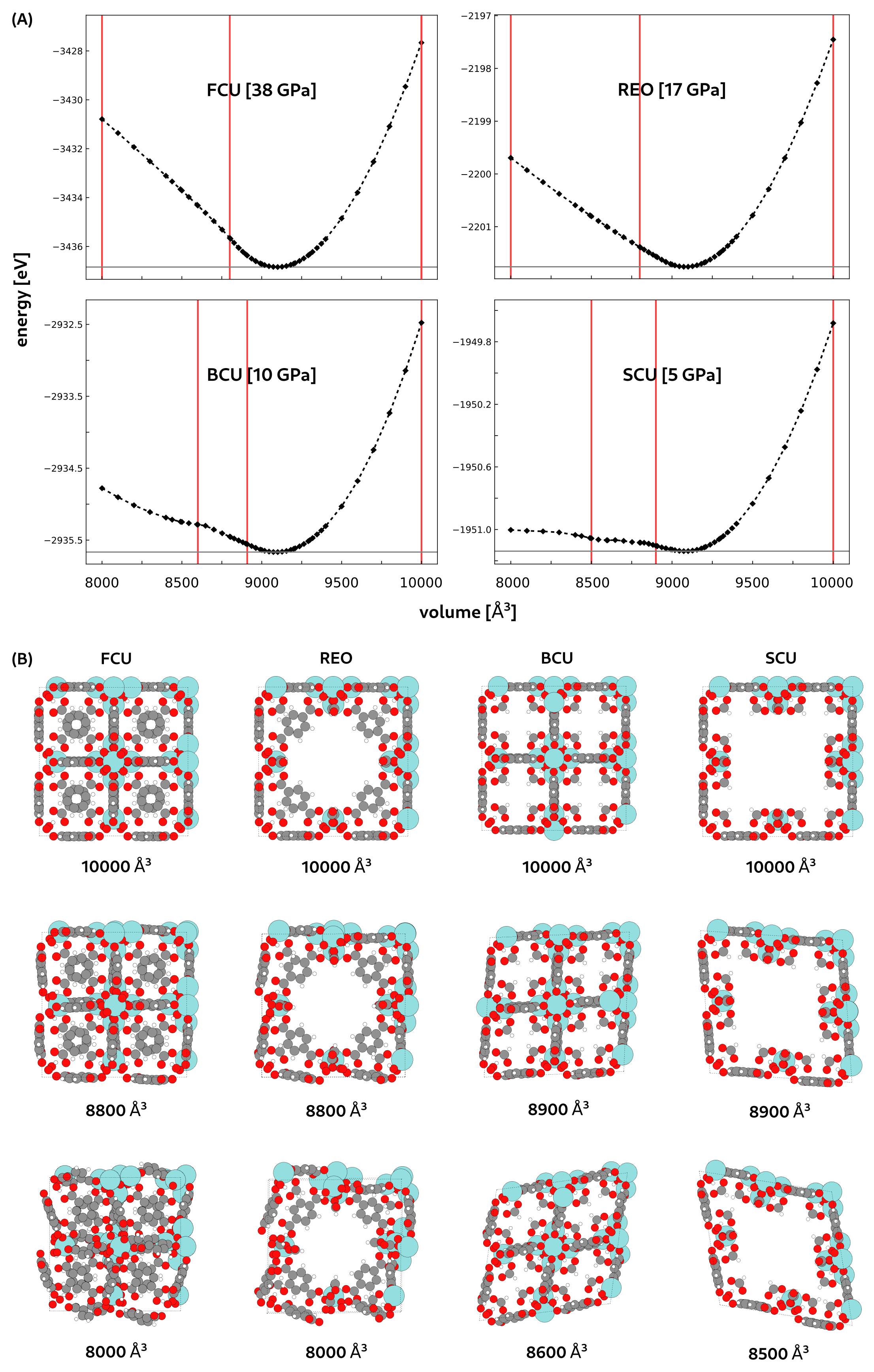}
\caption{Investigating deformation mechanisms of different UiO-66 topologies. Every EV curve samples the same volume array (A). Vertical red lines indicate volumes corresponding to the structural snapshots shown in (B).}
\label{si_fig_topologies}
\end{figure}

To explain the difference in mechanical behaviour between the fcu, bcu, reo and scu topological variants of UiO-66 ($S_\text{pr}$, $S_\text{bcu}$, $S_\text{reo}$ and $S_\text{scu}$), we investigate their structural evolution along an EV curve spanning a large volume range. This approach is preferred over a dynamical characterisation, as finite temperature phonons obfuscate deformation modes (more easily) observed at 0 K. Figure SI.\ref{si_fig_topologies}.A shows the corresponding EV profiles, computed at identical volume points for all four systems. In Figure SI.\ref{si_fig_topologies}.B, we provide snapshots of optimised structures for each topology, chosen at interesting volume points along the EV curve (see red markers). \\

The fcu and reo cells exhibit similar behaviour and will be discussed simultaneously. Starting at 10000 \AA$^3$, their elongated lattices are cubic and fully symmetrical. Cell compression is entirely accommodated by the shortening of covalent bonds. Symmetry is only broken around 8800 \AA$^3$, when bricks begin rotating and linkers twist out of their principal plane, although the cell shape remains mostly cubic. At smaller volumes, the effect is augmented and linkers lose their planar nature through buckling. This collective deformation mechanism involves every building block and requires significant energy, which we can deduce from the slope of the fcu and reo EV curve. The increased connectivity of the fcu topology explains its superiour resistance to applied pressures. \\

For the bcu and scu lattices, EV curves are qualitatively different. Above their equilibrium volume, both cells extend (or compress) anisotropically and maintain a tetragonal cell shape, because of asymmetric linker connectivity in different principal planes (see inset in Figure 7). At roughly 8900 \AA$^3$, a shearing deformation forms and demotes the bravais lattice to monoclinic. The shear does not directly exert stress on bricks or linkers and only reorients coordination bonds. As the volume decreases further, anisotropy increases and the structures skew more strongly, e.g., the angle between bcu cell vectors in the XY-plane decreases to $75 \degree$ at 8600 \AA$^3$. Instead of a smooth incline, the low-volume part of bcu and scu EV profiles show several kinks. These correspond to sudden jumps in shearing angle or slight reorientations of building blocks, when the optimisations branch between local PES minima. In Figure SI.\ref{si_fig_topologies}.B, the bcu and scu cells skew in opposite directions. This is simply a consequence of random symmetry breaking and not an actual property of the topology. \\

Note that the accuracy of \textbf{mlp}$^c_{\text{sup}}$ is not strictly tested for periodic structures with cell volumes below 8500 \AA$^3$. Nevertheless, given that its training set mainly consists of strongly out-of-equilibrium clusters, we are quite convinced of its extrapolation capabilities to smaller cells. Moreover, the framework deformations observed in Figure SI.\ref{si_fig_topologies}.B already start appearing around 8800-8900 \AA$^3$. While we cannot simply generalise these results to finite temperatures (300 K), the difference in deformation mechanisms - i.e., a limited reorientation of coordination bonds ($S_\text{bcu}$ and $S_\text{scu}$) versus a collective rotation of building blocks and contortion of ligands ($S_\text{pr}$, $S_\text{reo}$) - is a first clue explaining differences in mechanical behaviour for the chosen topologies.

\section{Overview of systems, datasets and MLPs} \label{si_overview}

\begin{figure}[]
\centering
\includegraphics[width=\linewidth]{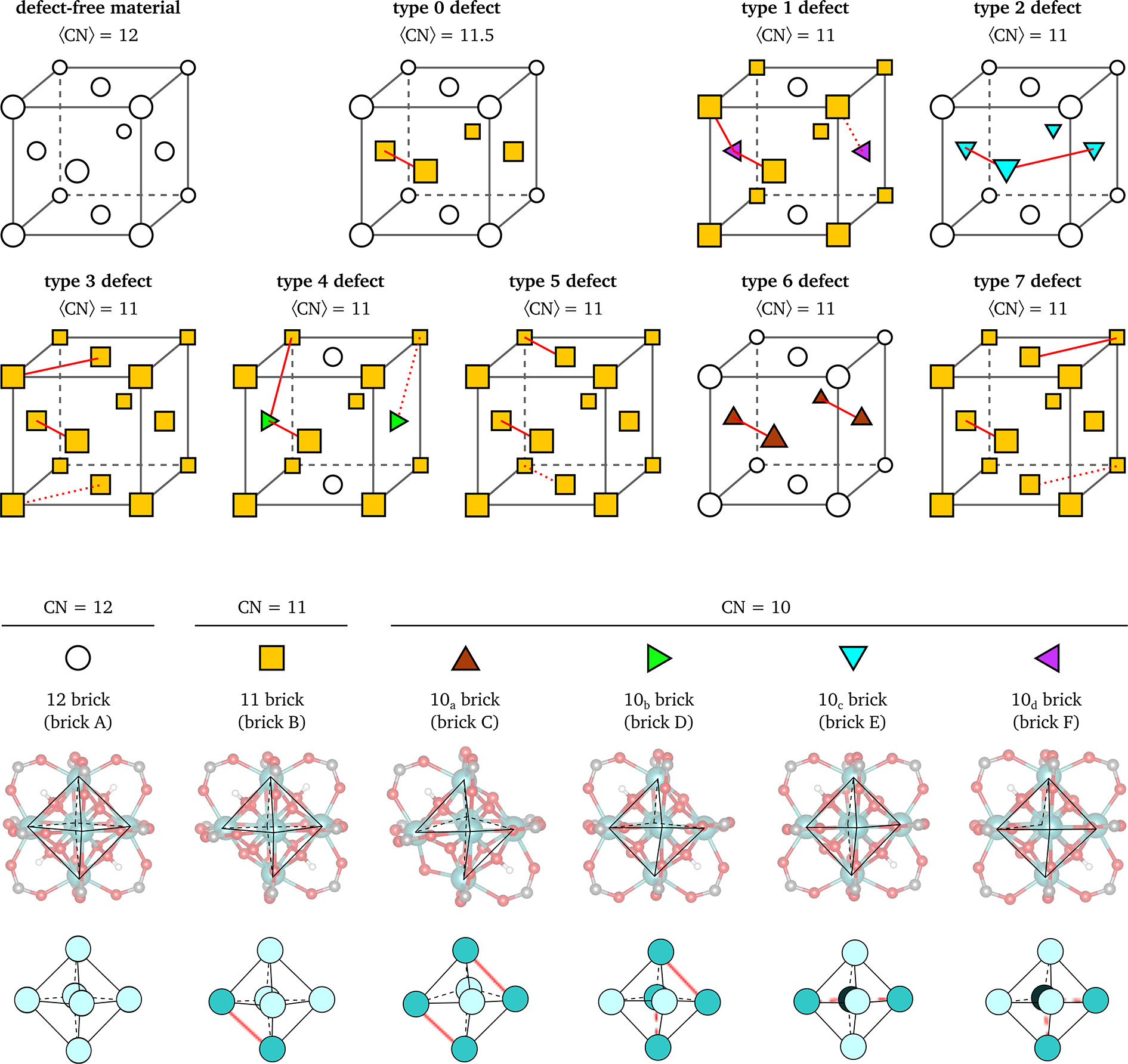}
\caption{Representations of UiO-66 unit cells: pristine ($S_\text{pr}$), single linker defect ($S_\text{ld}$) and all physically distinct combinations of a double linker defect ($S_\text{ld}^{1-7}$). Linker vacancies are indicated in red. Details of the Zr$_6$ octahedra are shown in the bottom pane, indicating the coordination number and missing ligands. The zirconium atoms are colour-coded based on their coordination number, and correspond, from light to dark, with a coordination number of 8, 7, and 6. Reproduced with permission from Ref. \cite{ref_rogge_thermodynamic_2016}. Copyright 2016, American Chemical Society.}
\label{si_fig_double_linker_defect}
\end{figure}

\begin{figure}[]
\centering
\includegraphics[width=\linewidth]{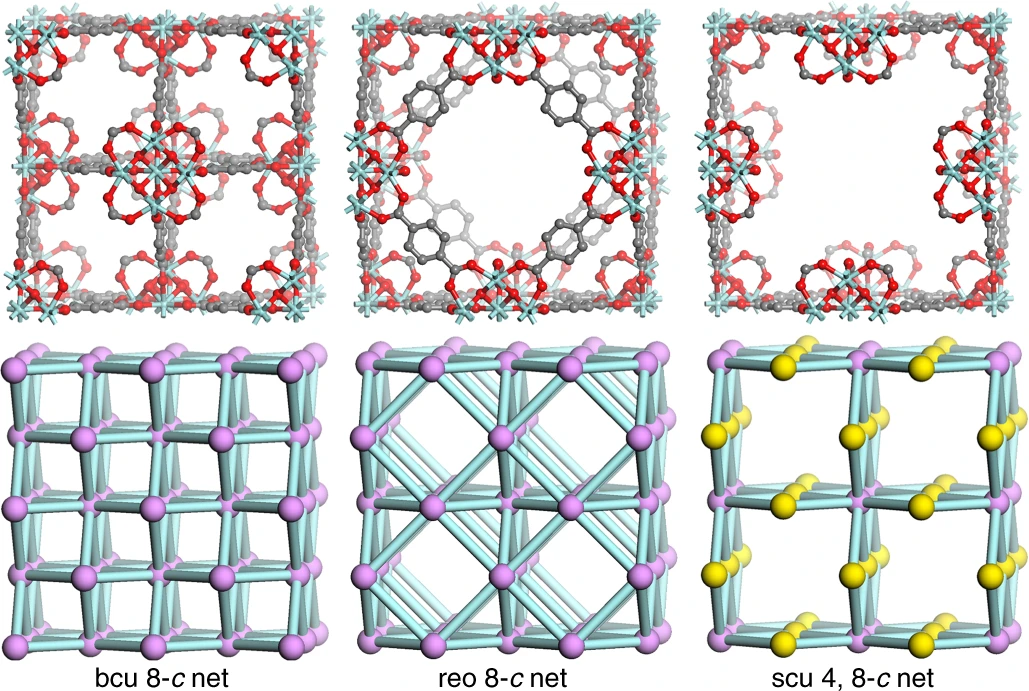}
\caption{Illustrations of various defective topologies in UiO-66. Top, crystallographic structural models. Bottom, corresponding topological representatives (2 × 2 × 2) of the 8-connected missing-linker defects (bcu net), 8-connected missing-cluster defects (reo net) and the 4,8-connected missing-cluster defects (scu net). Purple and yellow spheres indicate 8- and 4-connected nodes, respectively. Reproduced with permission from Ref. \cite{ref_liu_imaging_2019}. Copyright 2019, Springer Nature}
\label{si_fig_uio66_topologies}
\end{figure}

In this section, we provide an exhaustive overview of all systems, test datasets and MLPs that appear in Section 4.2 and Section 4.3 of the main text, see Table SI.\ref{si_table_systems}, Table SI.\ref{si_table_test_sets} and Table SI.\ref{si_table_models}. We also include some figures to accompany the defective unit cells introduced in Section 4.4 (see Figure SI.\ref{si_fig_double_linker_defect} and Figure SI.\ref{si_fig_uio66_topologies}). \\

\begin{table}[h]
\centering
\includegraphics[width=\linewidth]{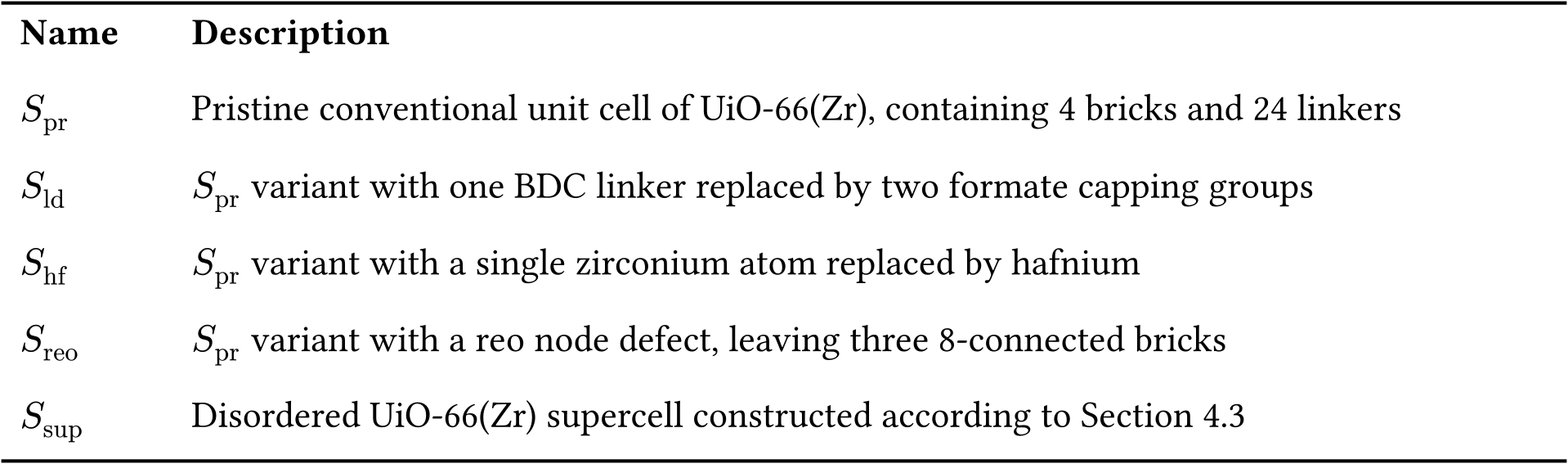}
\caption{All molecular systems used for training MLPs and extracting finite clusters. See Figure 5 and Figure 6 for (representative) visualisations.}
\label{si_table_systems}
\end{table}

\begin{table}[h]
\centering
\includegraphics[width=\linewidth]{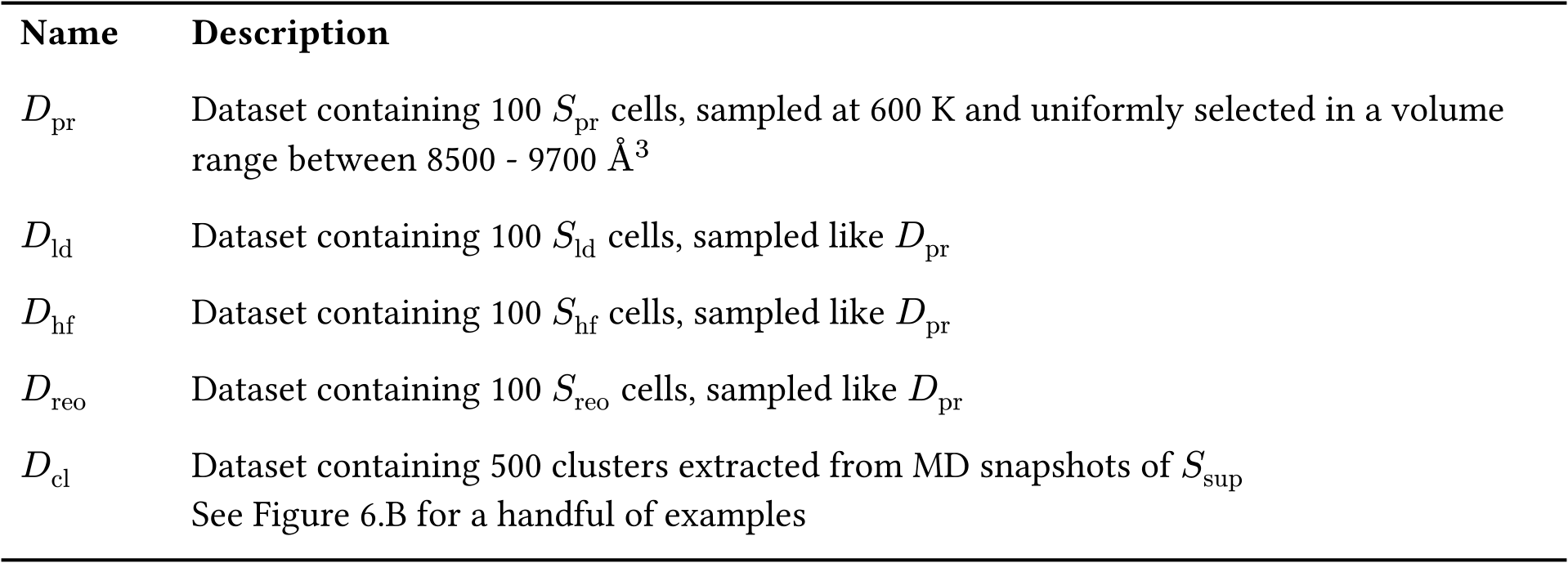}
\caption{All test sets used to benchmark MLP accuracy. Training sets - e.g., $D^T_\text{pr}$ - follow a similar naming scheme and are sampled analogously.}
\label{si_table_test_sets}
\end{table}

\begin{table}[h]
\centering
\includegraphics[width=\linewidth]{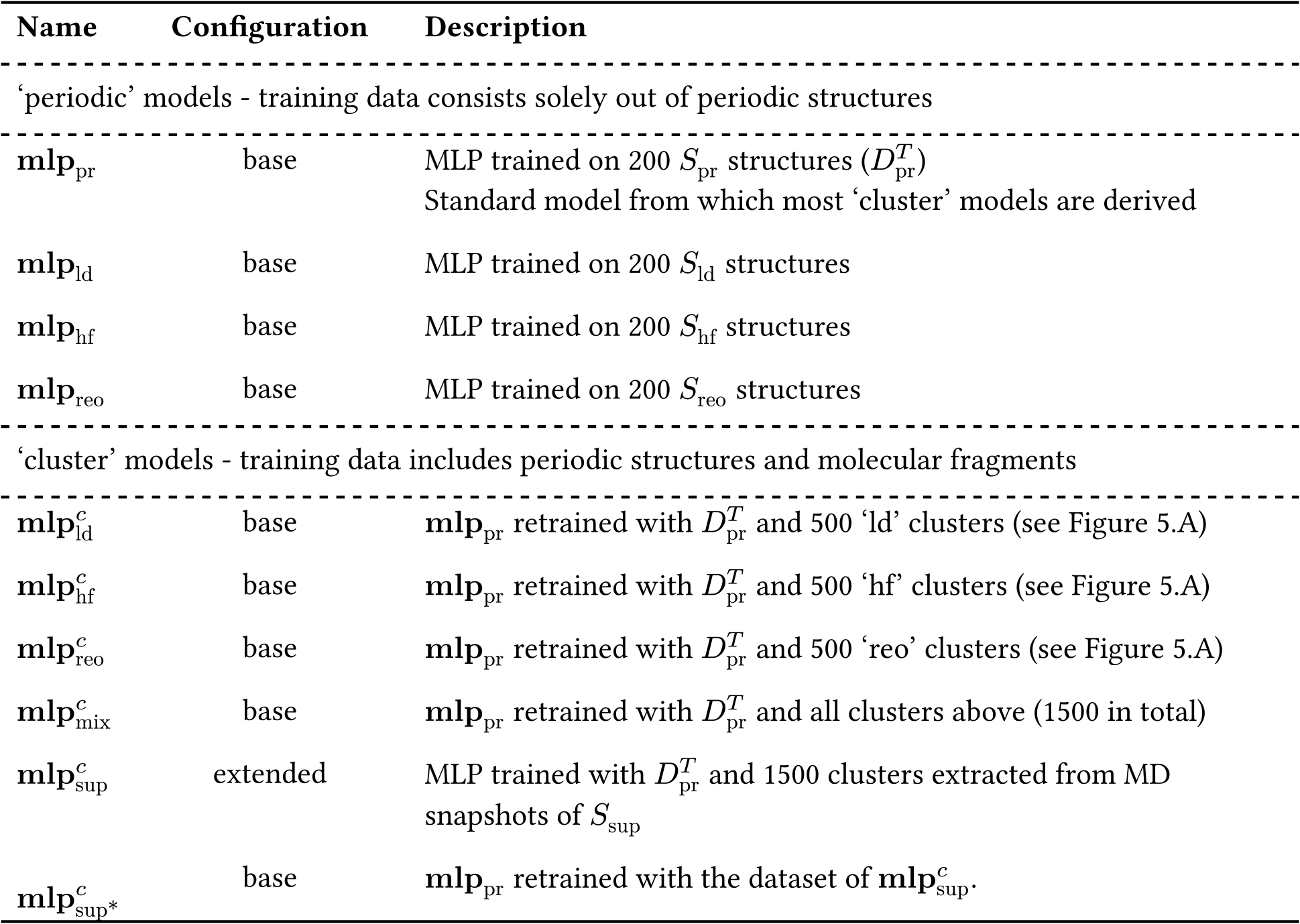}
\caption{An overview of every MLP trained in the main text, along with its hyperparameter configuration (see Section SI.\ref{si_nequip}) and a short description of its training dataset.}
\label{si_table_models}
\end{table}

\FloatBarrier

\section{MLP accuracy and test metrics} \label{si_metrics}

\begin{figure}
\centering
\includegraphics[width=\linewidth]{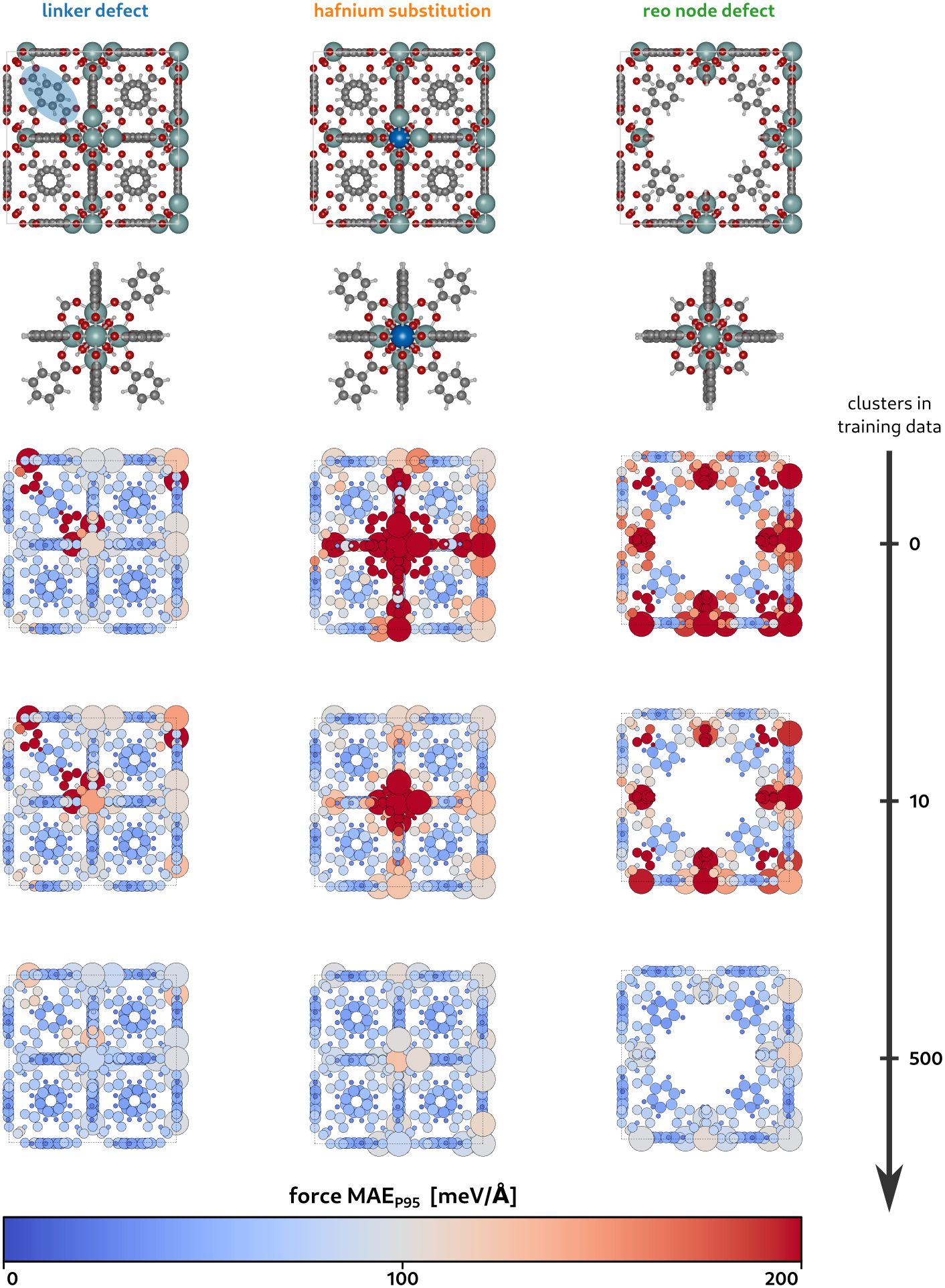}
\caption{Per-atom force $\text{MAE}_{P95}$ errors versus training data for cluster models along the $D_\text{ld}$, $D_\text{hf}$ and $D_\text{reo}$ learning curves discussed in Section 4.2. See also Figure 5.}
\label{si_fig_learning_curve}
\end{figure}

Figure SI.\ref{si_fig_learning_curve} shows supplementary per-atom model error plots for the learning curves discussed in Section 4.2, see also Figure 5.C. Numeric values are provided in Table SI.\ref{si_table_learning_curve}. Finally, we enumerate the evaluation accuracy of every model for all test sets in Table SI.\ref{si_table_metrics_full}. \\

\begin{table}[h]
\centering
\includegraphics[width=\linewidth]{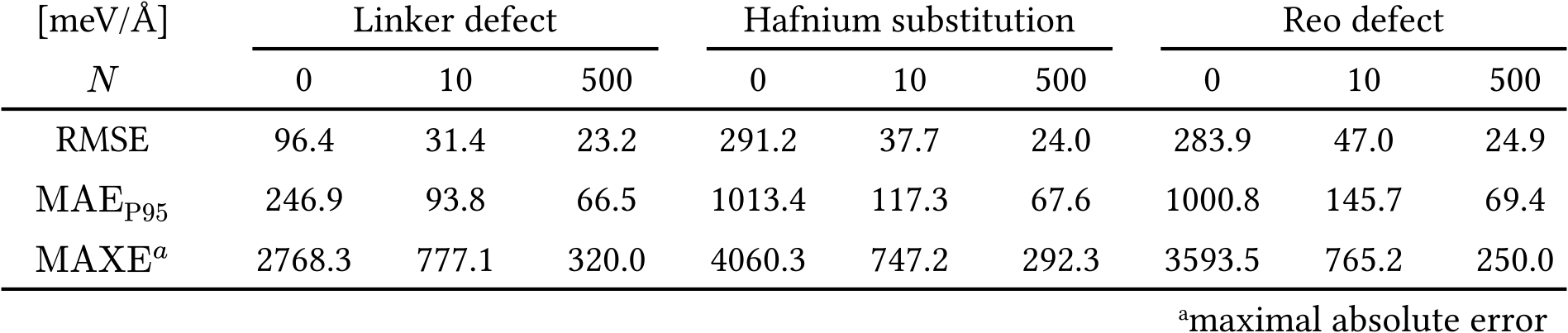}
\caption{Force error metrics for all cluster models of Figure SI.\ref{si_fig_learning_curve}, where $N$ represents the number of clusters of a given defect type added to $D^T_\text{pr}$.}
\label{si_table_learning_curve}
\end{table}

\begin{table}[h]
\centering
\includegraphics[width=\linewidth]{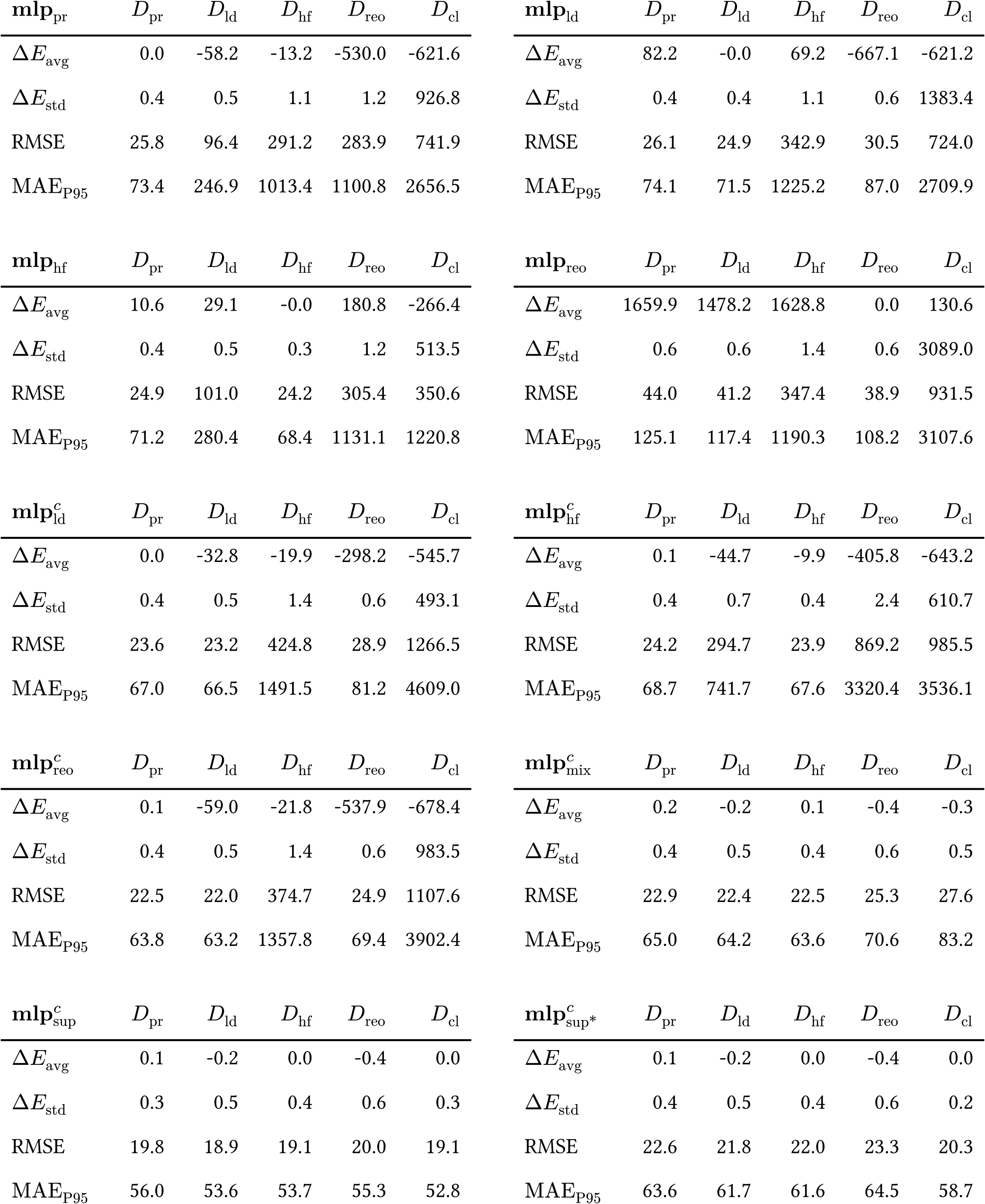}
\caption{Main error metrics for MLPs in Table SI.\ref{si_table_models} and test datasets in Table SI.\ref{si_table_test_sets}. $\Delta E_{\text{avg}}$ and $\Delta E_{\text{std}}$ values are expressed in meV/atom, RMSE and $\text{MAE}_{P95}$ in meV/\AA{}.}
\label{si_table_metrics_full}
\end{table}

\bibliography{refs}